**From Nuclei to Micro-Structure in Colloidal Crystallization: Investigating Intermediate Length Scales by Small Angle Laser Light Scattering**

Richard Beyer[1], Markus Franke[1], Hans Joachim Schöpe[1,2], Eckhard Bartsch[3] and Thomas Palberg[1]

[1]*Institut für Physik, Johannes Gutenberg Universität, D-55099 Mainz, Germany*

[2]*present address: Institut für angewandte Physik, Eberhard Karls Universität, D-72074 Tübingen, Germany*

[3]*Institut für Physikalische Chemie, Albert-Ludwigs-Universität, D-79104 Freiburg, Germany*

Hard sphere suspensions are well recognized model systems of statistical physics and soft condensed matter. We here investigate the temporal evolution of the immediate environment of nucleating and growing crystals and/or their global scale distribution using time resolved Small Angle Light Scattering (SALS). Simultaneously performed Bragg scattering (BS) measurements provide an accurate temporal gauging of the sequence of events. We apply this approach to studies of re-crystallization in several different shear molten hard sphere and attractive hard sphere samples with the focus being on the diversity of observable signal shapes and their change in time. We demonstrate that depending on the preparation conditions different processes occur on length scales larger than the structural scale which significantly influence both the crystallization kinetics and the final micro-structure. By careful analysis of the SALS signal evolution and by comparing different suggestions for small angle signal shapes to our data we can for most cases identify the processes leading to the observed signals. These include phase contrast form factor scattering from depletion zones during formation and overlap as well as during gelation, amplitude contrast form factor scattering by more transparent crystals, and structure factor scattering from late stage inter-crystallite ordering. The large variety of different small angle signals thus in principle contains valuable information complementary to that gained from Bragg scattering or microscopy. Our comparison, however, also shows that further refinement and adaptation of the theoretical expressions to the sample specific boundary conditions is desired for a quantitative kinetic analysis of micro-structural evolution.

## I. INTRODUCTION

Colloidal suspensions of spherical particles are a well recognized model system for condensed matter physics in general and for crystallization studies in particular [1, 2, 3, 4, 5, 6, 7, 8, 9, 10]. Over the last decade, considerable progress in understanding several important details of the crystallization process has been reported as reviewed in [11]. Experimental access exploits the convenient colloidal structural length scale, i.e. the scale of the inter-particle distance, which is in the nanometer to few micron range. Most studies on the mechanisms and kinetics of colloidal crystallization were therefore conducted employing either high resolution microscopy or Bragg scattering (BS) techniques. Microscopy yields information about individual particle positions and, thus, can reveal the involved mechanisms, while Bragg scattering observes large volumes with excellent statistics and therefore is best suited for kinetic studies. For instance, within both approaches a two step nucleation process has been established during which compactification and structural ordering occur on separate time scales [12, 13, 14, 15, 16, 17]. The two step process also influences the competition between crystallization and vitrification. In fact, in some cases a colloidal glass may be regarded as a dense packing of orientationally incompatible crystalline precursors [18, 19]. Further, solidification in mixtures [20, 21, 22, 23, 24, 25, 26] as well as heterogeneous nucleation were intensively studied [27, 28, 29].

In addition to the crystal structure, also the sample micro-structure, e.g. the arrangement, size and shape of crystallites, the defect pattern or a preferred orientation, are of great importance for the properties of solids. For colloidal solids, the micro-structure and its temporal evolution is well accessible by low resolution microscopy, such as Bragg microscopy [30, 23, 31, 32]. Further, suitable time resolved scattering methods may be employed to access the system evolution on intermediate length scales. This has been demonstrated already early by Ackerson and Schätzel [33] who used time resolved small angle light scattering (SALS) investigate crystallizing hard sphere suspensions. Typical length scales probed by SALS range from tens to hundreds of microns. As in microscopy [34], (scattering) contrast is provided either by fluctuations in refractive index or by local transmission changes due to scattering into the Bragg-regime. In their pioneering study the authors observed a ring-like scattering pattern at length scales corresponding to a few hundred lattice spacings. Angular averaging then yielded a peaked SALS signal. Ackerson and Schätzel interpreted their finding as evidence of the formation of a depletion zone about a growing crystallite of enhanced density as compared to the initial melt density. Together with the following series of papers, this was one of the first systematic studies on the crystallization kinetics of hard sphere suspensions [35, 36, 37]. Later, the interpretation of the SALS-signals was theoretically supported showing that they are generally expected for a local particle number density redistribution conserving average density over the volume of the fluctuation, but – depending on the refractive index variation and the growth mode of the fluctuation - may considerably differ in shape and intensity [38]. However, SALS studies remained rare and mostly date back more than a decade [39, 40, 41, 42]. In a more recent study, a combination of SALS with BS and microscopy was used to obtain insight in the unusual crystallization kinetics and micro-structural evolution of a short range attractive hard sphere polymer mixture prepared in the coexistence range close to the melting volume fraction [43, 44]. It nevertheless appears that in general the unique potential of SALS experiments to explicitly study the micro-structural length scale has not yet been fully exploited, mainly due to technical difficulties and challenges in signal interpretation.

One main intention of the present paper therefore is to draw direct attention to the diversity of SALS signals observable during crystallization, the other to demonstrate the possibilities to obtain valuable complementary information about the evolution of the sample structure on different scales by combining it with measurements in the BS regime. A key point is the identification of different possible origins for SALS signals during crystallization and we therefore compare to signal shapes suggested in literature. In fact, our present approach in data taking and interpretation draws strongly from SALS studies in other areas. For instance, depletion zone scattering seems to be responsible also for the formation of peaked signals during diffusion limited aggregation during the formation of isolated fractal aggregates in of sample coagulation [45, 46]. Also an additional length scale corresponding to steric cluster anti-correlation [47] may yield a peaked signal. Peaked signals are further well known from phase separation processes, e.g. spinodal decomposition [48, 49], where two homogeneous phases in sharp contact form a pattern with two characteristic length scales, namely the average linear dimension of a single phase and the repeat length scale. Their ratio and with it the peak shape, differ depending on whether the quench is critical or off-critical [49, 50]. Often, the shape of the signal stays constant in time and the scattering patterns superimpose, if scaled with the position and intensity of the maximum, $q_{MAX}(t)$ and $I_{MAX}(t)$. This has been termed dynamic scaling and implies a shape persistent self-similar scattering object which only grows in time, typically with a characteristic power law depending on the involved order parameter [51, 52, 53, 54]. However, not all peaked SALS signals show scaling [39, 45, 49] and a large fraction of SALS signals does not show a peak at all. For some cases, the latter can be related to an increased isothermal compressibility of the sample [55, 56]. Most commonly, inhomogeneities in scattering power yield a form factor scattering from either phase or amplitude contrast. Its power law decrease at larger scattering vector then corresponds to the dimensionality of the scattering objects [57]. Finally, a peaked signal may become hidden by averaging over a broad distribution of the characteristic length scales or if an additional scattering process gains in importance, e.g. from gelation in addition to spinodal decomposition [49].

Our study differs from previous crystallization studies employing SALS as we explicitly consider signals from samples taking known different solidification paths and thus deliberately allow for different signal origins. We therefore may ask how the SALS signal evolution depends on e.g. whether the sample shows fluid-crystal coexistence at equilibrium or has been prepared above the melting volume fraction or how the signal changes if weak or strong attractions of different range are introduced. In fact we demonstrate that for each of these cases characteristic SALS signals are observed which differ in both shape and temporal evolution.

Second, for each sample a *direct* comparison of simultaneously taken data from the two different scattering regimes was performed. To the best of our knowledge, this has not been done before. Relying on the meanwhile well established sequence of crystallization events on the structural length scale, this enables us to use our Bragg measurements to create a reference time scale for the simultaneous SALS experiment.

Third, by comparison to the simultaneously taken BS data and to models of the SALS signal we can for the first time discriminate differing origins for similar looking SALS signals as well as different kinds of SALS signals corresponding to different SALS objects present during different stages of crystallization. Refraining from previous attempts to interpret these data as evidencing processes mainly on the structural length scale, we gain new qualitative information about the sample micro-structure. Any quantitative characterization of micro-structure evolution will, however, need additional systematic measurements and a refined modelling adapted to the specific circumstances.

Finally, some of the experiments were performed on a machine originally reported by Heymann et al. [58]. This machine was well suited for the micro-structure evolution during the coarsening stage, because it was sufficient to identify the end of the main crystallization stage. For all other studies we used a refined version with diode arrays replaced by CCD arrays, which allowed to obtain a reliable temporal gauging with precision Bragg data also during the early stages of crystallization.

The remainder of this paper is organized as follows. In Section II, we i) introduce our samples and their preparation followed by ii) a short discussion of the experimental set-up detecting SALS and Bragg scattering (BS) signals simultaneously. For a more detailed description the interested reader is referred to Appendix A. Section III presents i) the procedure of temporal gauging with the example of a HS sample prepared at coexistence and ii) a compares signal shapes and evolution in BS and SALS for all five investigated samples to obtain some basic information on their crystallization kinetics. In Section IV, we discuss on the observed SALS signal shapes and their origin. After i) some general remarks we compare our findings to theoretical suggestions for ii) samples in the fully crystalline state, iii) samples prepared at coexistence and iv) samples with late stage signals. We also v) address the question, of characteristic SALS signal types for different crystallization scenarios. We finish with some short conclusions.

## II. EXPERIMENTAL

### i) Samples

HS crystals come in several close packed structures (face centred cubic, fcc; hexagonal close packed, hcp; or randomly stacked, rhcp) which show a slow ageing towards the equilibrium fcc structure [59]. The location of the first order freezing transition of monodisperse HS, however, is well defined and depends only on volume fraction, $\Phi$ [1, 3, 60]. Polydispersity may alter the exact positioning of the freezing and melting line, but it is only for large polydispersities that fractionation is encountered [61, 62]. With some non-adsorbing polymer added, the system acquires an attractive contribution to the potential of mean force, with the range of the attraction depending on the polymer size and its strength on the polymer concentration. This so-called depletion interaction considerably changes both the phase behaviour and the crystallization scenario [6, 7, 8, 9, 10, 11, 24, 63, 64, 65, 66]. Fig. 1a displays a theoretical prediction for the phase behaviour of monodisperse hard spheres with added polymer as derived from computer simulations on a system with size ratio $\xi = r_g/a = 0.1$, where $r_g$ and $a$ are the radius of gyration of the polymer coils and the radius of the (swollen) colloidal particles, respectively [67]. Shown are the phase boundaries in the $\eta_{res} - \Phi$ plane, i. e. the polymer-reservoir packing fraction – particle volume fraction plane. With increasing polymer concentration the coexistence region widens in a characteristic fashion with a pronounced step-like feature in the volume fraction of the coexisting solid at $\eta_{res} \approx 0.05$. Fig. 1a also illustrates the choice of preparation conditions made for the present study in order to cover a broad range of different crystallization scenarios.

Open diamonds represent the locations of the melt volume fractions, $\Phi_0$, of our samples HS1 to AHS2, while stars represent their measured crystal volume fractions, $\Phi_{Crystal}$. Comparison of the respective locations indicates the amount of crystal compression, which, as expected, is large for AHS2 with $\Phi_0 < \Phi_M$. Crystal compression is only small for the other samples prepared either in the fully crystalline state ($\Phi_0 > \Phi_M$, HS2), or just below the melting volume fraction (HS1), or at an $\eta_{res}$ yielding only a marginal widening of the coexistence region (AHS1, AHS3). We note, that also our third attractive sample (AHS3) of low polymer content, $\eta_{res} \approx 0.077$, and $\Phi_0 < \Phi_M$ showed only mild crystal compression [43, 44, 65]. It differs from AHS1 and AHS2 by its much smaller range of attraction (c.f. Tab. I).

While for a given volume fraction the equilibrium crystal structure of colloidal HS is fixed, the crystal micro-structure can be varied by several other factors. For instance, it is strongly influenced by gravity introducing changes in morphology, gradients of particle concentration and/or stratification [68, 69, 70, 71, 72, 73, 74]. To minimize such effects, we carefully buoyancy-matched our systems. We used 1:30 cross-linked Polystyrene (PS) micro-gel spheres which have been synthesized by the group of E. Bartsch and are representative for a large class of similar particle systems [24, 28, 43, 44, 75, 76]. Tests of their hardness of interaction employing oscillatory rheological measurements following [77] return coefficients of an inverse power potential. For 1:10 (1:50) cross-linked systems, we typically find n = 90-100 (n = 40-50) at a swelling ratio, $V_{swollen}/V_{dry}$, of 2.2-2.8 (5-6) [78, 79]. From a simple interpolation of n versus swelling ratio we here expect n = 70-80 for the present particles with $V_{swollen}/V_{dry} \approx 3$.

For a detailed description of synthesis and sample preparation we refer to [24, 28]. Our stock for HS1 to AHS2 is a freeze dried powder of 1:30 cross-linked particles with dry radius of 276 nm and a polydispersity index of $\sigma = 0.072$ as obtained from TEM measurements. Particles were re-suspended in 1-Ethylnaphtalene (1EN) to achieve a good match of both mass density and index of refraction ($n_{D,PS} = 1.599$, $n_{D,1EN} = 1.606$, $\rho_{PS} = 1.05$ g/cm$^3$, $\rho_{1EN} = 1.008$ g/cm$^3$). The particles swell in this good solvent over a time span of two weeks to a final hydrodynamic radius of about a = 0.395 μm. Note that swelling further improves the matching. This is readily seen in Fig. 1b. Even a month after solidification, one hardly notes the meniscus in AHS1 at coexistence, indicating that most of the fluid is still contained in the grain boundary regions and not separated by crystallite settling. During the crystallization experiments, no sedimentation effects were discernible at all.

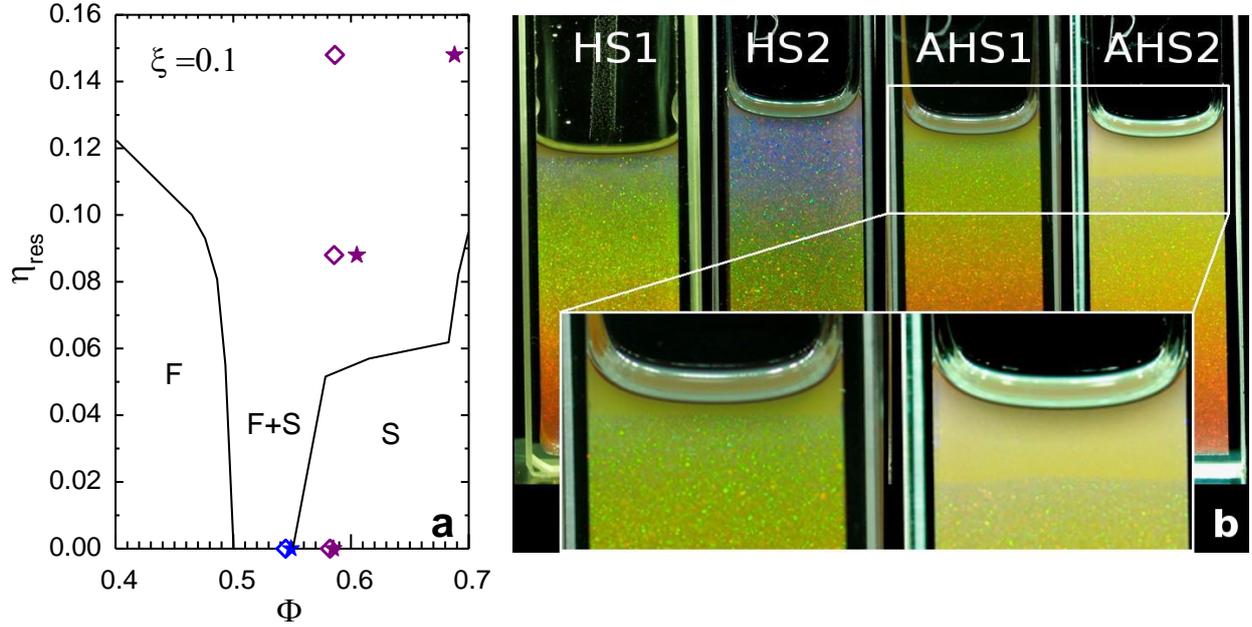

*Fig. 1: Sample characteristics. a) Locations of investigated samples HS1 to AHS2 in the $\Phi$ - $\eta_{res}$ – plane of the phase diagram. Solid lines are redrawn from [67] and denote the boundaries between the fluid (F), coexistence (F+S) and solid (S) regions of the phase diagram for monodisperse HS with $\xi = r_g / a = 0.1$. Open diamonds denote initial volume fractions of the meta-stable melt, stars denote final crystal phase volume fractions; blue: HS1; violet(from bottom to top): HS2, AHS1, AHS2. Note the significant crystal compression occurring at large polymer reservoir volume fraction. b) Buoyancy match and crystal settling in samples HS1 to AHS2 as seen in cells of 10 mm width four weeks after last homogenization. Only for AHS2 an extended zone of fluid ordered supernatant is visible.*

Samples investigated are compiled in Tab. I. For the pure HS samples, phase boundaries (in terms of weight fraction) were determined using the method of Paulin and Ackerson [80] and then mapped onto the theoretical freezing line (in terms of volume fractions) for polydisperse HS with $\sigma = 0.072$ ($\Phi_F = 0{,}521$, $\Phi_M = 0{,}572$ [81]). HS1 and HS2 represent HS with start volume fractions $\Phi_0$, below and above the experimental melting volume fraction $\Phi_M = 0.548$, respectively. The final volume fractions of the crystal phase were calculated from the position of the $(111)_{FCC}$ Bragg reflection in q-space, $q_{MAX}$, as: $\Phi_{crystal}(t) = 2(q_{MAX} a)^3 / (9\sqrt{3} \pi^2)$ [82]. Further, $q = 4\pi n/\lambda_0 \sin(\theta/2)$ is the modulus of the scattering vector with laser wave length $\lambda_0$, refractive index of the suspension $n$, and scattering angle $\theta$ (see also below).

To HS2 different amounts of non-adsorbing polystyrene of molecular weight of $M_w = 819000$ g/mol (*PSS Polymer Standards Service GmbH, Mainz, Germany*) were added to obtain the attractive samples AHS1 and AHS2. Employing the light scattering based data of [83], the manufacturer determined $M_W$ was converted to a radius of gyration of $r_g = (40.6 \pm 1.0)$ nm yielding a size ratio of $\xi = r_g / a = 0.102$. The polymer reservoir concentrations $\eta_{res}$ were calculated following [63]. The stock for AHS3 contains 1:10 crosslinked PS-micro-gel particles of dry radius 300 nm and a polydispersity index of $\sigma = 0.06$. The swollen radius is a = 380 nm and, from synthesis, about 2% w/w of PS oligomers with a molecular weight of $M_w \approx 1074$ g/mol are present in the suspension. This results in a size ratio of $\xi = 0.008$.

**Table I: Samples and sample properties**. Samples HS1 to AHS2 are 1:30 cross-linked Polystyrene (PS) micro-gel spheres suspended in buoyancy and refractive index matching 1-Ethylnaphtalene (1EN). a: swollen radius of particles; σ: polydispersity index; $\Phi_0$: start volume fraction of the meta-stable melt; $\Phi_{Crystal}$: final volume fraction in the crystal phase; $\eta_{res}$: reservoir volume fraction of polymer, ξ: size ratio, data for AHS3 are reproduced from [43, 44]

| Sample/Lab Code | a / nm | σ | $\Phi_0$ | $\Phi_{crystal}$ | $\eta_{res}$ | ξ |
|---|---|---|---|---|---|---|
| *HS1/KS15* | *395* | *0.072* | *0.5445* | *0.549* | *0* | *-* |
| *HS2/ K9RP* | *395* | *0.072* | *0.5822* | *0.585* | *0* | *-* |
| *AHS1/ K9RP* | *395* | *0.072* | *0.5856* | *0.605* | *0.0822* | *0.102* |
| *AHS2/ K9RP* | *395* | *0.072* | *0.5863* | *0.688* | *0.148* | *0.102* |
| *AHS3/Mn4* | *380* | *0.06* | *0.54* | *0.540* | *0.077* | *0.008* |

### ii) Light scattering instrument and data processing

Schätzel and Ackerson's early instrument consisted simply of a screen placed behind the laser-illuminated sample and observed with a video camera [33, 35, 36, 37]. Later, a second camera monitoring turbidity allowed characterizing the scaling of the SALS peak [39, 40]. The first instrument for simultaneous BS and SALS [58] used two photo diode arrays rotating about the optical axis to capture the complete 2D-scattering pattern in both regimes. Azimuthal averaging significantly improves signal statistics. Further it leaves the delicate samples mechanically undisturbed, which is particularly important for studying the still fluid crystal environment during solidification. We therefore use this approach for our experiment. However, we employ eight CCD array detectors, which considerably enhance angular resolution, as was first demonstrated by Francis et al. [84] for the BS-regime.

Fig. 2 shows a schematic side view of our set-up. Illuminating optics, sample stage, BS and SALS detection systems are mounted to a sturdy rack placed on a vibration isolating table. Both detector units rotate about a common axis defining the system's optical axis (dashed line). The ball bearing bedded rails for each detector set are mechanically coupled and synchronously driven by a computer controlled stepper motor [*AMIS, Germany*] with suitable home built gearing. For a more detailed description the reader is referred to Appendix A.

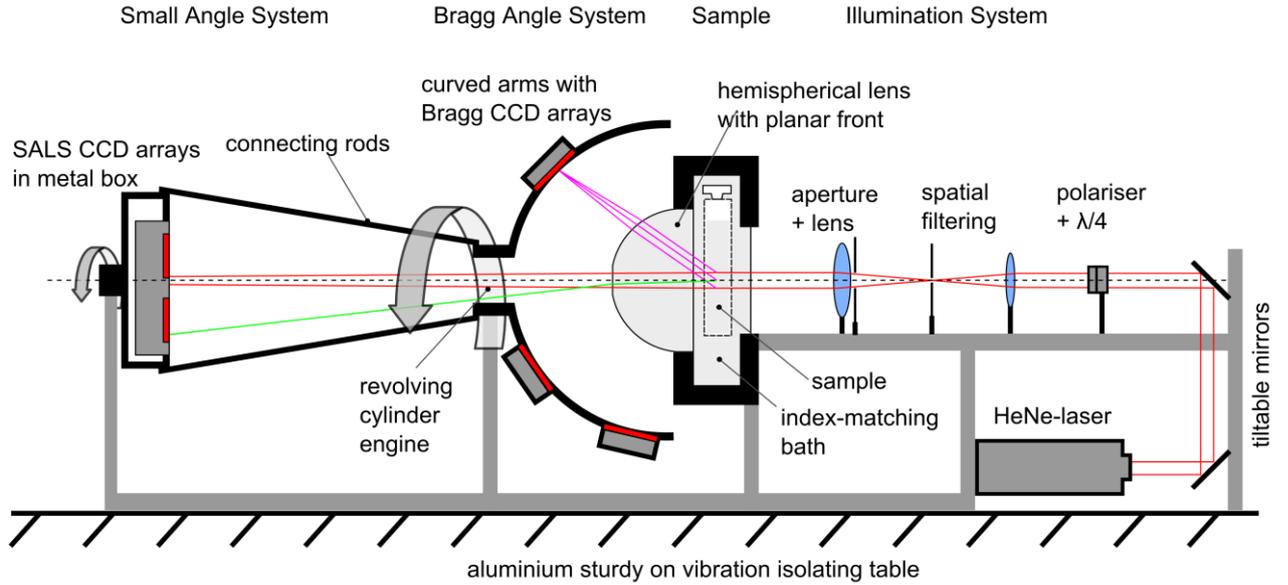

*Fig. 2: Schematic side view of the combined BS and SALS experiment with components indicated. There are six CCD detectors on four rotating curved rails to detect the BS signal and two for the SALS signal. Rotation of both detection systems is mechanically coupled and driven by a single computer controlled stepper motor.*

Time resolved static light scattering experiments on crystallizing samples follow an established protocol. Prior to the experiment, samples are kept in a shear molten homogenized state on a tumbling wheel for at least 24 hours. Taking the sample off the tumbler and mounting it to the instrument takes about one minute. Start of the instrument defines t ≡ 0. Scattering signals from both regimes are monitored at increasing time intervals for up to 40 hours. The raw SALS signal is corrected for background contributions, this time by subtracting the first measurement $I(q,0)$: $I_{SALS}(q,t) = I(q,t) − I(q,0)$ [36, 42]. This neglects any significant contribution of scattering from the melt to the SALS signal which would disappear during crystallization. At low $q$, the melt structure factor for homogeneous samples, $S(q)$, approaches a value $S(0) = \rho \kappa k_B T$, where $\rho$ is the particle number density, $\kappa$ denotes the isothermal compressibility and $k_B T$ is the thermal energy. For pure HS systems $S(q \to 0)$ is independent of $q$, very low and hardly different for melt and crystal phase. The difference between melt and crystal scattering may therefore be neglected in isolating density inhomogeneities due to an evolving micro-structure. For AHS, the situation is somewhat different. Again, over the investigated q-range, $S(q)$ is fairly independent of $q$, but it may attain a much larger value than for the crystalline phase [55, 56]. The changing melt scattering contribution may therefore become non negligible. We take up this point again below in the discussion of AHS3. In the BS regime, scattering contributions from the melt and the crystalline phase superimpose, but now display markedly different q-dependencies. However, at t = 0, only the meta-stable melt contributes. Following [82], we therefore isolate the crystal scattering contribution by subtracting the first measurement, $I(q,0)$, from the recorded intensity: $I_{Xtal}(q,t) = I(q,t) − \beta(t)I(q,0)$, with $\beta(t)$ being a scaling factor representing the decreasing fraction of melt.

### III.  SIMULTANEOUS MEASUREMENT OF BRAGG AND SMALL ANGLE SCATTERING

#### i)  Discrimination and Calibration of time scales

The main difference of the present approach as compared to previous BS & SALS measurements is to use the precision BS data to construct a temporal gauge for the SALS data. From this calibration of crystallization processes at the structural length scale we will obtain a basis for the interpretation of the SALS signal. This has to be done for each different situation separately and the principle is now demonstrated using the example of a HS sample at coexistence. The scattering patterns of HS1 are displayed for both the BS and the SALS regime in Figs. 3 a and b, respectively. Both figures are divided into panels (A) to (C) each comprising a characteristic stage of the crystallization process. We note that this discrimination is solely based on the features directly visible in the scattering patterns.

In Fig. 3a we see the typical temporal evolution of the crystalline BS intensity distribution for a random stacking of hexagonal close packed layers (r-HCP) [85]. Note, that only the two relevant q-ranges are displayed which show on the left the fine structure of the principle peak and on the right the higher order peaks. As compared to the original instrument the present angular resolution was increased from $\Delta q \approx 0.9$ µm$^{-1}$ to $\Delta q = 1.41 \cdot 10^{-3}$ µm$^{-1}$ per data point with a low noise level and a large dynamic range. Such fine resolution has been shown to allow for a peak by peak analysis of the BS data with a high temporal resolution and good statistics [19, 28, 65, 82, 84, 86]. For early times (A) a fuzzy, broad main peak develops, caused by the simultaneous formation of bulk precursors and wall crystals [12]. Later in (B), the signal obtains a defined substructure. One discriminates a first $(100)_{HCP}$-, a main $(111)_{FCC}$-peak and a third peak in between which stems from crystals nucleated with their $(111)_{FCC}$ lattice plane at the container walls (heterogeneously nucleated wall crystals [28, 41]). At the same time the higher order $(220)_{FCC}$ and $(311)_{FCC}$ peaks become visible as well. The first occurrence of $(220)_{FCC}$ indicates a well ordered 3D FCC structure without any wall crystal contribution. It is discriminated from the following stage shown in (B') in which a $(101)_{HCP}$-peak emerges while the wall crystal peak stops evolving further and is submerged between the $(100)_{HCP}$ and the main $(111)_{FCC}$ reflection. In fact, a preferred and strong intensity increase of the $(111)_{FCC}$-peak and of the higher order peaks is observed, while the intensities of both the $(100)_{HCP}$ and the $(101)_{HCP}$-peak increase much slower. This indicates a relative reduction in the fraction of ABA stacking sequence in favour of ABCA stacking. Further, one notes a pronounced shift of the peak position towards lower q-values corresponding to a decrease in particle density going from initially highly compressed precursors to growing crystallites [82, 86]. Finally, (C), the Bragg scattered intensity nearly saturates with a small, constant amount of equilibrated fluid remaining.

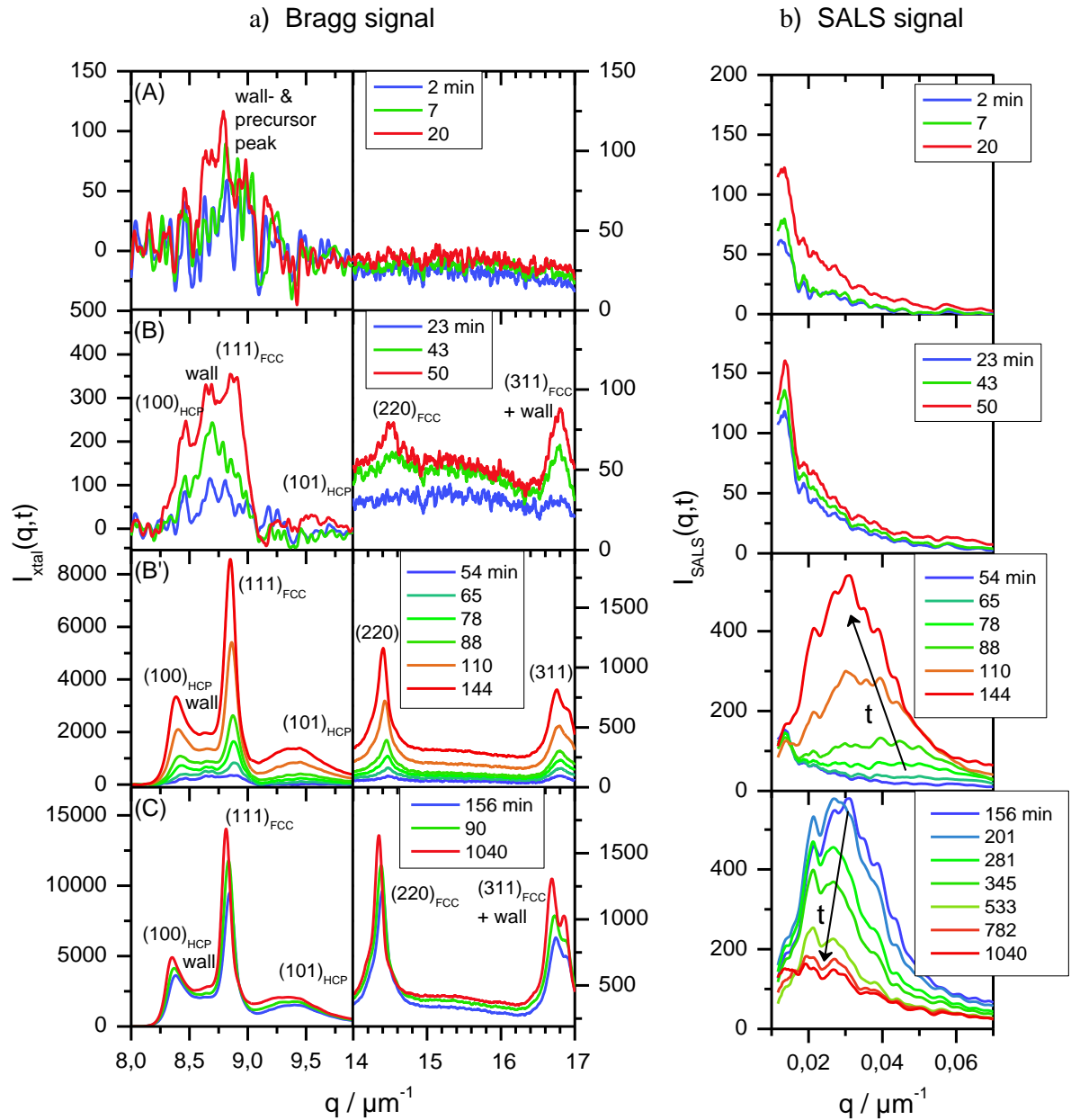

*Fig. 3: Evolution of the scattered intensity with time of sample HS1 with time colour coded in each panel from small (blue) to large times (red). (a) In the BS regime, the final scattered intensity from crystals $I_{Xtal}(q,t)$ shows five well defined crystalline peaks which are attributed to r-HCP crystals in the bulk and at the wall. Four stages of their evolution in time may be discriminated. (A): precursor stage with the appearance of amorphous precursors in the bulk and at the wall; (B): transition stage from induction to crystallization with first appearance of bulk crystallites and growth of wall crystals; (B'): main crystallization stage with conversion by crystal nucleation and growth in the bulk and (C): ripening stage. (b) The SALS-signal shows two distinct signal types, a narrow peak-like structure at small q and a broad peak-like structure that first grows then declines with its maximum position steadily shifting towards lower q (as indicated by the arrows). Again, the three time regimes (A) to (C) can be well discriminated.*

In Fig. 3b we display the simultaneously measured results from the SALS regime. Also here we draw from an enhanced angular resolution as compared to the original instrument but further exploit the shift of the detection range towards lower q. This allows to demonstrate that the shape of the SALS signal changes significantly throughout the crystallization process from (A) to (C). Already for very early times (A) the SALS signal shows a clear increase in intensity in the q-range $q < 0.06$ μm$^{-1}$. This signal comprises of a pronounced peak-like structure at small $q < 0.015$ μm$^{-1}$ and a more gentle increase of the intensity for $q > 0.015$ μm$^{-1}$. This signal shape is retained throughout stages (A) and (B). The increase in overall intensity is fast throughout (A) but significantly slower during stage (B). An additional, initially rather broad, peaked feature emerges during (B') at large q $(0.018 \leq q \leq 0.06)$ μm$^{-1}$ which quickly gains intensity. It simultaneously sharpens to form a pronounced SALS peak in which the small-q feature seen during (A) and (B) is successively submerged. The position of its maximum continuously shifts towards lower q as indicated by the arrow. It reaches $q = 0.02$ μm$^{-1}$ at the end of the (B') stage of the BS data. Interestingly, the SALS signal further changes during (C), while the evolution of the BS signal has already come to a halt. The SALS peak position continues to shift towards lower q, but now the overall intensity decreases.

To compare the temporal evolution in both regimes, we start by looking at the integrated intensities I(t). For the BS regime we calculate the overall intensity $I_{BS}(t)$ of the group of the first three Bragg peaks ($(100)_{HCP}$, $(111)_{FCC}$ and $(101)_{HCP}$) within the q-range of (8.1 to 10.5) μm$^{-1}$. For the SALS signal, $I_{SALS}(t)$, we integrated over the complete accessible SALS q-range. Fig. 5a compares the integrated intensities for both regimes in a linear (Fig. 5a) and in a double logarithmic fashion (Fig. 5b). After an initial lag time of some 25 min the BS signal gains rapidly in strength to saturate after some 200 min. Quite differently, the SALS signal shows an initially weak increase, followed by a rapid increase and a decrease after some 200 min. Although the signals differ qualitatively, the temporal evolution of the two data sets is clearly correlated. Further, the BS signal neatly corresponds to the two step nucleation scenario also observed on other HS systems [36, 37]. This novel crystallization scheme was originally suggested in [12], later confirmed in extensive computer simulations [13, 14, 15] and by confocal microscopy [16]. As of now, it seems well established that crystallization of HS follows a scenario involving several steps [11, 28]. In this scenario, amorphous precursors are formed first, which then convert to crystals which grow and further increase in number by parallel nucleation from the remaining melt, until finally the sample enters a coarsening regime, in which the amount of crystalline material stays constant but the microstructure develops further through ripening processes.

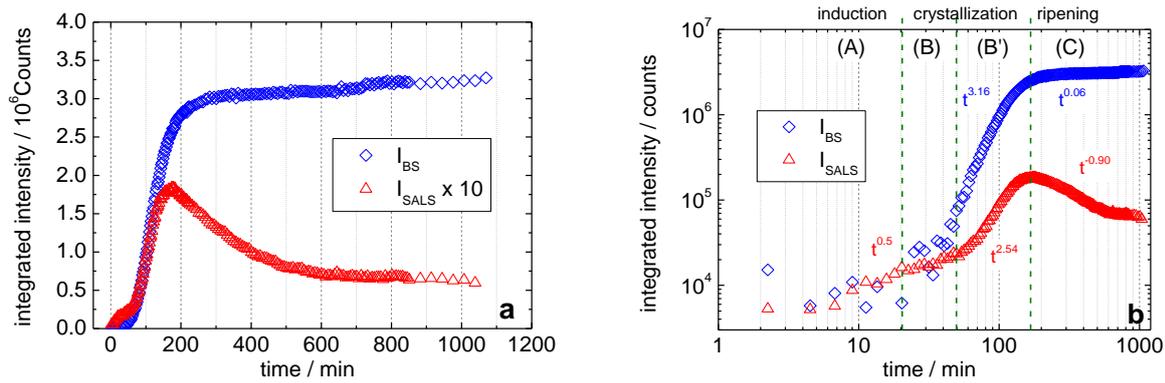

*Fig. 4: Temporal evolution of integrated scattered intensities of the (111) BS peak group (diamonds) and the SALS (triangles) regime. a) Linear representation with the SALS data multiplied by a factor of 10 for clarity; b) double logarithmic representation. Where possible, the corresponding power laws for the temporal evolution are indicated. The stages (A) to (C) are indicated and separated by vertical dashed lines. $I_{BS}$ is constant at a low, noisy level during (A), shows a strong increase during (B) and (B') and a plateau during (C). $I_{SALS}$ was obtained from integrating over the complete SALS q-range. It shows a power law like growth during (A) and (B), a steeper increase during (B') and a decrease during (C). The integrated intensity peaks at the transition from the conversion to the ripening stage.*

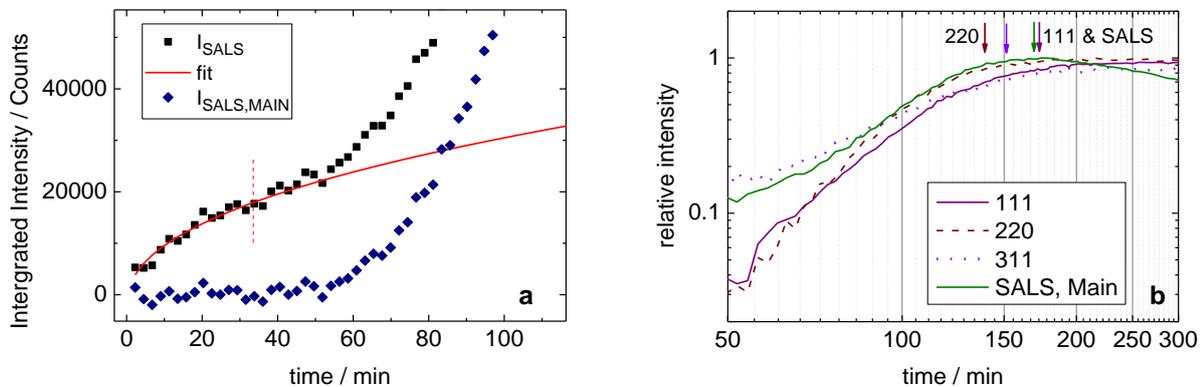

*Fig. 5: Determination of characteristic times. a) Onset of the main SALS signal. Squares: SALS intensities integrated over the complete q-range. Solid line: square root fit to the data for $t \leq 35$ min (indicated by dashed vertical line). Diamonds: experimental data after subtraction of fit; The main signal emerges from the noise level at $t \approx 50$ min. b) Transition from the main crystallization process to the coarsening stage. Data normalized to their maximum values.*

A closer look at the data reveals that the start of stage (B) at about 25 min is clearly seen in the BS regime from the first appearance of higher order peaks. An onset of stage (B') is less visible due to the initially small signal to noise ratio for the $(101)_{HCP}$-peak and the continuous increase of intensity for all other peaks in Fig. 3a. Further also the integrated BS intensities in Fig. 4b show a large scatter. The situation is somewhat more defined for the SALS signal. Fig. 3a is suggestive of an early rise in integrated intensity which tends to slow and is superseeded by the strong rise in intensity during the main crystallization stage. In Fig. 4 one notes that the SALS data can be well approximated by a $t^{0.5}$ power law increase at early times. The development of the SALS signal is shown enlarged for early times in Fig. 5a. Note that there is no change of the temporal development in this signal at 25 min, rather, the pronounced increase occurs at some later time. Referring to the evolution of signal shape seen in Fig. 3b, we assume that this signal is composed of two superimposed signals. Several fit functions were tried for the early time data. We found that only a square root function decsribed the SALS data taken for t ≤ 35 min well. To discriminate the onset of the second signal, we then subtract the fit function from the data. The thus isolated main SALS signal emerges from the noise level at t ≈ 50min (Fig. 5a). The same characteristic time is found, if we fit two power law functions to the SALS data of Fig. 4b for the stages before and after the kink and determine their crossing point. Both fixes the onset of a stage (B') to t ≈ 50min. The choice of the fit range induces a systematic uncertainty of about four percent. Most important, this new stage is seen in the SALS signal only and therefore denotes a micro-structural process independent of the crystallization processes occurring on the structural length scale.

Fig. 5b shows a comparison of integrated intensities with data normalized to their maximum values for the time window corresponding to the transition from stage (B, B') to stage (C). The transition is rather broad. Further, the early SALS data contain a dominant contribution from the initial small-q signal. While the data neatly coincide across the transition period, a quantification of cross-over times appears to be difficult. Still, the crossing points of power law fits to the data for the different BS reflections and the SALS maximum more or less coincide with a variation between 140 min (220) and 170 min (111 & SALS). Such a variation is significant, since the systematic errors of the fit procedure are about ten percent. This shows that i) the transition from stage (B) to stage (C) is not exactly abrupt. This supports previous observations that, growth and coarsening processes may occur simultaneously [86, 87]. ii) crossover times are not exactly coinciding. This is indeed expectable since SALS and different BS peaks monitor different processes. In BS both the total amount of crystal (dominantly within the low-q peaks) and the crystal quality (dominantly for the higher order peaks) contribute. For SALS we monitor the micro-structural development, thus e.g. an overlap of the depletion zones formed around growing crystallites may occur independent of the termination of crystal growth. iii) For HS1 and other pure HS samples in the upper part of the coexistence region, the start of the decrease of the SALS signal coincides with the crossover time determined from the low-q BS peaks as indicated by the arrows in Fig. 5b. This points to a coupling of structural and micro-structural processes for HS1 but may not necessarily be the case for other samples (see below).

Following [28], we use the BS data to order and label the sequence of processes on the structural length scale. We label the first stage (A) as "induction" and its scattering features are attributed to precursor formation. The next stages (B) and (B') together are labelled "crystallization" and interpreted as the conversion of precursors and melt to crystalline phase by further nucleation and crystal growth. The final stage (C) is labelled "ripening" and its spectral features are explained by coarsening processes in the crystallized sample. We stress that with the improved BS angular resolution this gauging can be performed for any crystallizing sample. The discrimination between (B) and (B'), however, is hardly possible on the basis of the BS data alone, but may be performed utilizing the SALS data. Thus one may define (B) beginning with the first appearance of crystalline ordered regions as visible by higher order Bragg reflections and (B') starting with the appearance of the peaked SALS signal due to some micro-structural process.

### ii) Comparison of the evolution in signal strength and shape for differently prepared samples

The data shown above for HS1 are typical for pure HS in the upper part of the coexistence region. Samples prepared under different conditions show different signals. In any case, however, the temporal gauging based on the simultaneously measured BS data can be performed and employed in their analysis. The characteristic times as extracted from the BS data gauging are displayed in Tab. II. An overview on the SALS signal shapes and the evolution of the integrated intensities in the two scattering regimes is presented in Fig. 6 for all investigated samples. Integration was carried out over the complete q-range for the background corrected SALS signals and for the low-q peak group, i.e. the q-range covering the $(111)_{FCC}$, the $(100)_{HCP}$ and the $(101)_{HCP}$ reflections in the background corrected BS signal [82]. In Fig. 7 the integrated intensities are normalized to their maximum values for comparison. The SALS signals differ in shape and temporal development for the different samples. Further, I(q,t) is seen to differ for BS and SALS. The two regimes therefore contain complementary information on different aspects of the solidification process. A unique interpretation of all different SALS signals e.g. in terms of a crystallite size seems very difficult [88]. Any specific interpretation will need a suitable model for the SALS signal origin and additional information from either the BS data and/or microscopy. This will be attempted in the next section.

| Sample | $t_{ind}$ / min | $t_{cross}$ / min |
| --- | --- | --- |
| **HS1** | 24 | 170 |
| **HS2** | 9 | 56 |
| **AHS1** | 160 | 520 |
| **AHS2** | 650 | 2600 |
| **AHS3 (This work / [42, 43])** | - / - | 52 / 9 |

**Table II: Characteristic times.** $t_{ind}$: induction time marking the onset of stage (B); $t_{cross}$: crossover time marking the transition from the crystallization (B/B') to the ripening stage (C) as obtained from the integrated intensity saturation of the low-q group of reflections. The systematic uncertainty in this procedure ranges between ten percent for HS1 and some fifty percent for AHS2. For sample AHS3 no (A) stage is discernible. Further, we also quote the result obtained for AHS3 in [42, 43] from the evaluation of the (220)-reflection which is not affected by wall crystal growth and twinning issues. The value of 9 minutes thus

reflects the end of crystal growth, while the value of 52min reflects the saturation in crystal quality. Note the significant shift of time scales towards larger values occurring with increasing polymer concentration for samples HS2 to AHS2 and with lower degree of meta-stability for samples HS2 to HS1.

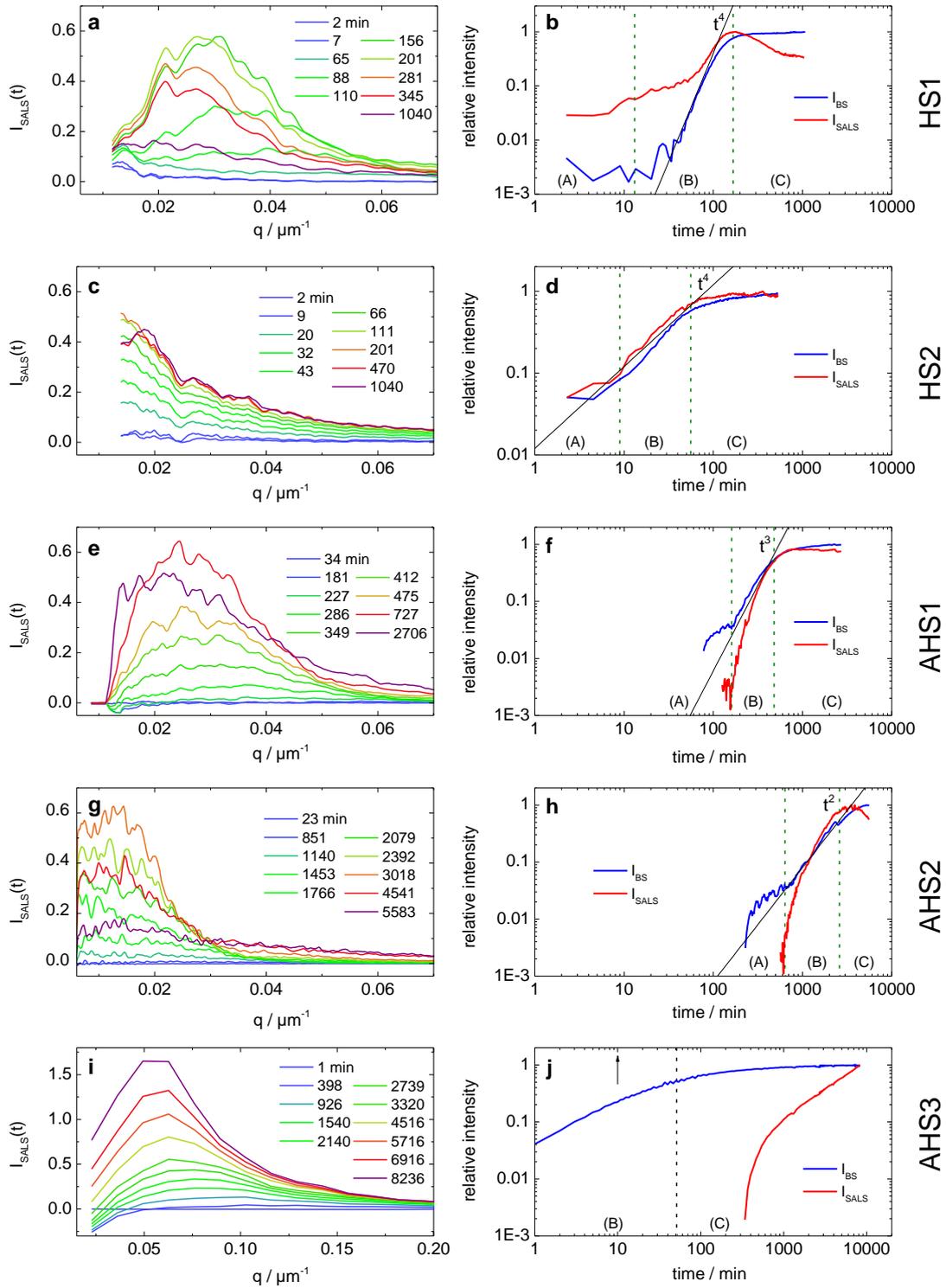

*Fig. 6: Comparison of samples HS1 to AHS3. Note the increased q-space resolution in a, c, e and g as compared to i, which was taken with the previous setup. Left column: evolution of the SALS signal from early times (blue) to late times (violet) as indicated. HS1 and AHS1 show peaked signals which first gain then*

*loose intensity. HS2 shows a signal for which the intensity decreases monotonously with q and increases monotonously with time. AHS2 shows a signal of complex and time dependent shape with a non monotonous intensity development. AHS3 again shows a peaked signal, however of continuously increasing intensity. Note that here the peak appears at a larger q-scale than for HS1 and AHS1. Right column: comparison of the temporal development of BS and SALS integrated intensities, normalized to their maximum values. Solid black lines are guides to the eye of power laws with the corresponding exponents indicated. The BS signal intensity shows a sigmoidal shape for HS1 and HS2. It shows a step-like behaviour for AHS1 and AHS2. Note that a significant slowing of the overall dynamics is observed for these three samples of same Φ when the strength of attraction is increased. For AHS3 the earliest times are not recorded and only the upper part of a presumably sigmoidal shape is visible. The SALS signal intensity for HS1 shows two steps. For HS2 it is strictly parallel to the BS signal intensity. For AHS1 the rapid intensity increase during (B) is followed by a slow decrease during (C). For AHS2 the decrease was followed for only four days, but it appears to be more pronounced. For AHS3 an increased SALS intensity becomes discernible only long after the beginning of coarsening (C). This sample was investigated still using the previous machine of Heymann et al. showing a much lower angular resolution. The analysis in [43, 44] revealed a power law increase in the integrated intensity with a coarsening exponent α ≈ 1/3 corresponding to a conserved order parameter.*

## IV. DISCUSSION OF POSSIBLE ORIGINS OF SALS SIGNALS

### i) General considerations

All observed scattering ultimately relies on the scattering contrast of individual colloidal particles and the distribution of the colloid density fluctuations. Descriptions based on this have been derived covering both the BS and the SALS regime [89]. In the following we use a description treating SALS and BS on the same footing, but separately [90]. This allows to account for different scattering mechanisms and detection schemes in the two regimes. As usual, for spherical scatterers at any given q in each regime, the scattering intensity, I(q), factorizes according to $I(q) = I_0(q) N b^2(0) P(q) S(q) H(q)$ [5, 11]. Here, $I_0(q)$ is proportional to the illuminating intensity $I_0$ modified by the details of the experimental set-up. This apparatus function differs for the two experiments but is constant in time. N is the number of scattering objects in the scattering volume, $b^2(0)$ is the single object scattering cross section as determined for q = 0, $P(q) = b^2(q)/b^2(0)$ is the form factor of a scattering object, S(q) is the static structure factor reflecting the spatial correlation of scattering objects and H(q) is a shape factor characterizing a spatial region of homogeneous structure. Note, that this general description can be applied for both BS (where the scattering objects are colloidal particles) and SALS (where the scattering objects are medium scale fluctuations of some kind). For non-spherical objects like e.g. for aligning rods or preferentially oriented anisometric crystals P(q) and S(q) do not factorize.

For BS, the measured intensity is background corrected and the scattering of crystals is isolated by subtracting the fluid signal [82]. N is then given by the number of colloidal particles in favourably oriented crystallites and increases in time. The structure factor S(q) in this regime is a constant series of delta functions and is convoluted by the changing crystallite form factor H(q) allowing an interpretation of the BS peak widths in terms of crystallite sizes [91]. For SALS, N is given by the number of large scale objects providing a time dependent $b_{SALS}^2(0,t)$ and $P_{SALS}(q,t)$ through their time dependent particle density distribution and/or BS scattering cross section. Note that for any given SALS scattering object $b_{SALS}^2(0)$ may originate from both phase and amplitude contrast. Several choices are possible for $P_{SALS}(q,t)$. One may e.g. employ Furukawa's function, which reads: $P(Q) = (1 + \delta/2) Q^2 / (\delta/2 + Q^{\delta+2})$ with $Q = q / q_{MAX}(t)$ and the free parameter $\delta$ which depends both on the dimensionality, d, of the scattering object and on whether the quench is critical ($\delta = 2d$) or off-critical ($\delta = d+1$) [49, 57]. Derber et al. derived a function corresponding to a growing crystal surrounded by a growing depletion zone under the assumption of overall mass conservation [38]. For signals due to pure amplitude contrast between Bragg scattering crystals and surrounding melt one may employ Rayleigh-Debye-Gans form factors for spheres or disks. Finally, The SALS structure factor is 1, unless there evolves an ordered micro-structure and there is no H(q)-analogue for the SALS case except for the shape of the illuminated volume. Fig. 7 shows a comparison of exemplary SALS signal shapes expected for different SALS objects. Peaked signals are expected for spinodal decomposition type conversion and for crystals with depletion zone, monotonously decaying signals for solid spheres in a homogeneous matrix. Next, we will first discuss the unpeaked signal of HS2 and then the peaked signals.

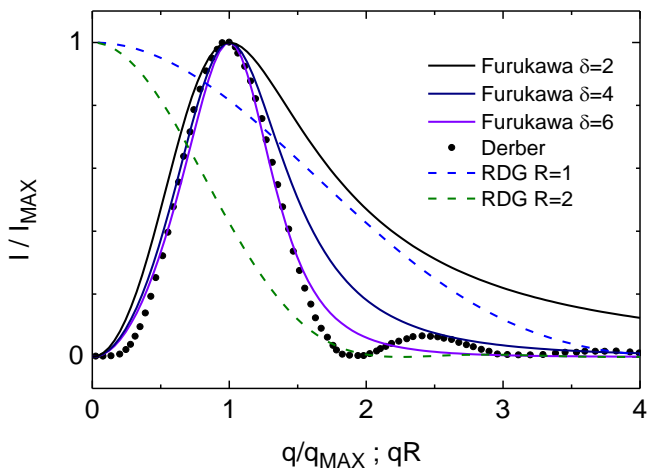

*Fig. 7: Comparison of different theoretical SALS signal shapes. Furukawa function for different dimensionalities of self similar SALS objects with sharp scattering contrast boundaries. Dimensionality increases from top to bottom (solid lines); Derber's function for a solid sphere surrounded by a diffusional depletion zone calculated with mass conservation [38] (dotted line). These are plotted versus $Q = q/q_{MAX}$.*

*Also shown are Rayleigh-Debye-Gans form factors for solid spheres of different radii, R, (dashed lines) plotted versus qR. R increases from top to bottom.*

### ii) Fully crystalline pure HS samples (HS2)

The SALS signal shape of HS2 remains unpeaked throughout all stages. Its intensity decreases steadily with increasing *q*. Moreover, it increases in time in remarkably close agreement with the BS intensity. From the analysis of the BS signal we further know, that the integrated intensities of both the 111 group of reflections and the 220 peak which is void of wall crystal contributions increase roughly with $t^4$, indicating reaction controlled growth and simultaneous nucleation at constant rate density [88]. Since under such conditions no extended depletion zones evolve [38], we do not expect the SALS signal to be peaked. Rather we suggest, that it is due to form factor scattering at the growing crystals. Form factors have their maximum intensity at forward scattering, for q = 0. Averaging over differently sized crystals washes out the minima of P(q) present for each individual crystal (c.f. Fig. 7). Increase in crystal size with time will increase the scattering intensity.

In principle, the SALS contrast may result from both phase and amplitude fluctuations. The former is well known from conventional static light scattering [1]. In SALS we consider the altered refractive index within a crystal as compared to the surrounding melt leading to the form factor of a sphere. To obtain the signal for the complete sample we assume the superposition of the individual form factors of all growing 3D crystallites. Amplitude contrast is well known from bright field microscopy but has rarely been considered in connection with light scattering on crystallizing samples. Interestingly, a randomly oriented crystal will in general not fulfill the Bragg condition and therefore will scatter much less than the surrounding fluid. Only if it meets the Bragg condition, it may – depending on its thickness – scatter much more efficiently [92]. For a suspension of randomly oriented optically thin crystals one therefore expects the transmission to increase during solidification. This in fact has been exploited to study the nucleation rate density of charged colloidal suspensions [93]. SALS signals in addition contain information about the scattering object's shape and dimensionality equivalent to the information contained in a B/W-image of the sample obtained in bright field microscopy. For our case of randomly oriented crystals we therefore expect a superposition of N individual form factors of 2D discs. To discriminate between the two possible contrast mechanisms, we display in Fig. 8 the SALS signals of HS2 normalized to one at q = 0.05 $\mu m^{-1}$ in a double-log plot. We compare them to Rayleigh-Debye-Gans form factors [57, 89] for i) solid 3D spheres of radii R = 80 µm and R = 60 µm (red circles and squares, respectively) and ii) to the form factor of a circular 2D disc with R = 80 µm (black diamonds). Clearly, the calculated curves for the 3D objects fail to reproduce the data in the large-q regime, while the disc form factor describes the data remarkably well. This clearly shows that for HS2 amplitude contrast is responsible for the formation of the SALS signal.

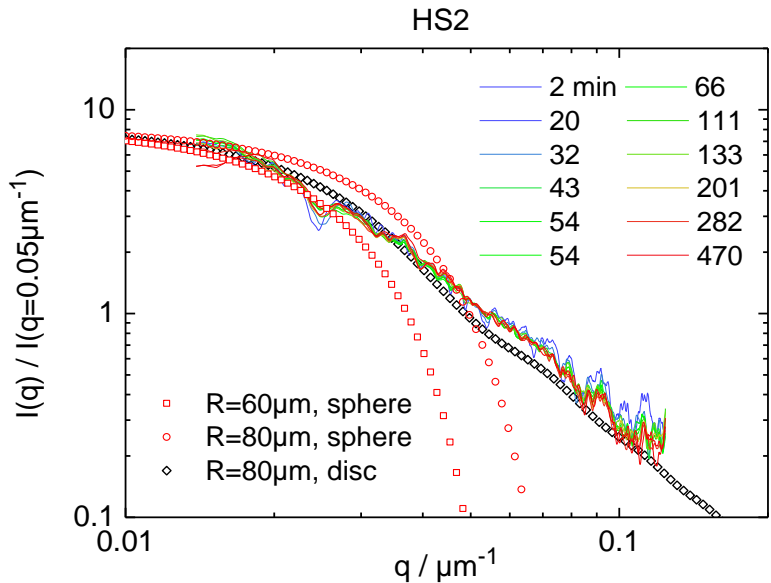

*Fig. 8: double-log plot of the SALS data of HS2. The superimposed data show a large-q power-law decay with an exponent close to 2. Red squares and circles show sphere form factors calculated within Rayleigh-Debye-Gans approximation for radii of R = 60µm and R = 80µm, respectively. Black diamonds denote the form factor for opaque circular discs of R = 80µm. The agreement to the latter is convincing and we conclude that our SALS signal in HS2 corresponds to scattering from disk shaped objects of diameter 2R and thickness t << 2R observed on scales 2R > 1/q > t [57].*

Further, also the early stage SALS signal observed for sample HS1 displays a $q^{-2}$ decay for 0.03 µm$^{-1}$ < q < 0.1 µm$^{-1}$. In fact, such a signal is often observed for samples prepared at equilibrium coexistence. It is much rarer for samples above equilibrium coexistence. Its origin remains unsettled. However, following the discussion above we feel that it might result from local transmittance changes at wall nucleated crystals appearing immediately after the quench at the hard container walls. These precede any bulk crystal formation and their fraction dominates the total crystallinity at early times. By contrast, above coexistence, wall crystals are rarely growing to a well visible extension and thus should not yield a significant SALS contribution.

Finally, the noise on the signals in Fig. 8 appears to be correlated. We checked that this effect of a correlated sinusoidal substructure is present in practically all spectra of a single crystallization experiment, but the noise is not correlated between different crystallization runs on one sample, nor between runs on different samples. This may indicate a systematic origin either introduced by subtracting the first run or due to Fresnel diffraction within the samples or sample cells. Since, however, the relative amplitude of this noise is acceptably small, we neglect it in the further discussion.

### iii) Samples at coexistence (HS1, AHS1 and AHS2)

The development of the SALS signal of samples prepared at equilibrium coexistence is very similar. In all cases a broad, peaked feature gains in intensity and shifts slightly towards the lower q. Samples however differ in the correlation of the SALS peak development and the development of signals in the BS regime, and further also in the position of the peak. AHS3 develops the peak only way into stage (C) and is therefore discussed separately in the next section. HS1, AHS1 and AHS2 show an increase in peak intensity during stage (B). Later, during stage (C), the intensity decreases again. HS1 and AHS1 show a pronounced peak with $(0.02 < q_{MAX} < 0.04)$ µm$^{-1}$, while the maximum for AHS2 is rather shallow and found at $q_{MAX} < 0.02$, µm$^{-1}$ much closer to the origin.

To determine the positions and heights of the maxima, we fitted Furukawa functions [49, 50] to the scattering signals of Fig. 6a, e and g and assign $I_{MAX}(t)$ and $q_{MAX}(t)$ to the maxima of the fit curves. In Fig. 9 we re-plot the SALS signals of samples HS1 and AHS1 (Fig. 6a and e) scaled by $I_{MAX}(t)$ and $q_{MAX}(t)$ as obtained from the fitting procedures with $Q = q/q_{MAX}$. When the intensity becomes significantly larger than the noise level during (B') the signals superimpose in the peak region and at larger $Q > 2.4$. For the range $1.2 < Q < 2.4$ the experimental curves show a small but significant and systematic increase in scaled intensity with time (Fig. 9a and b). During (C) the intensity in the high-q-tail $Q > 2.4$ starts to increase as well for both samples (Fig. 9c and d). To be more specific, only the slope increases, the curves still approaches zero for larger Q.

We compare the scaled signals to Furukawa functions [50] for 3D-SALS objects after critical quench ($\delta = 6$), 2D-SALS objects after critical or 3D SALS objects after off-critical quench ($\delta = 4$) and 1D-SALS objects after off-critical quench ($\delta = 2$). We further compare to Derber´s form factor of a crystal with homogeneous density $\Phi_M$ and radius R and a diffusive depletion zone starting at R with $\Phi_M$ and relaxing outward towards $\Phi_0$ (dashed line) [38]. We note that all function types describe the peak region fairly well.

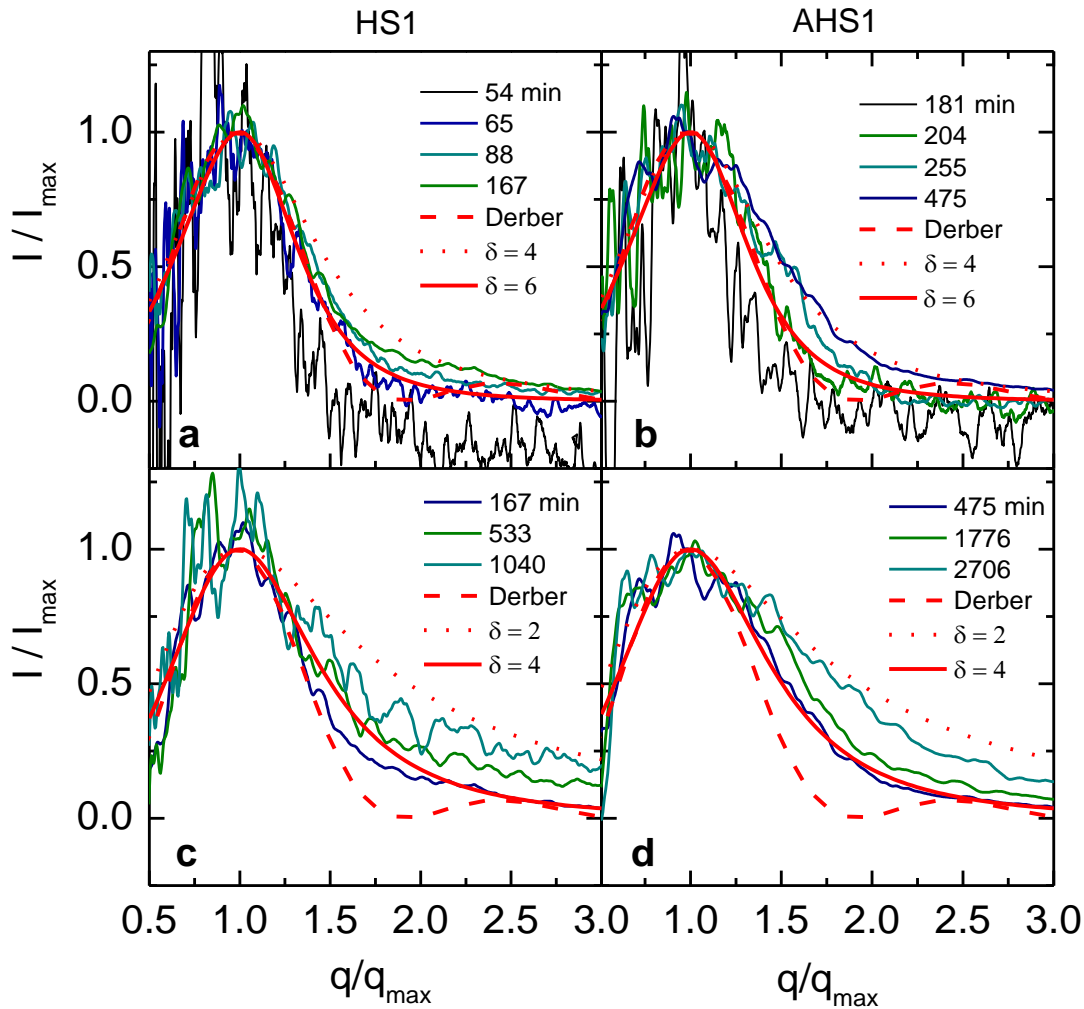

*Fig. 9: Rescaled SALS data of samples HS1 (left column) and AHS1 (right column) taken at times as indicated. Upper row: stage (B); lower row stage (C). The data tend to collapse to a single curve during the main crystallization stage (B), but not during coarsening (C). We further compare the signal shapes to scaling functions proposed in the literature: Furukawa function for $\delta = 2$, 4 and 6 as indicated and Derber's form factor of a spherical crystallite with a diffusive depletion zone.*

Not for very early times throughout stage (B), but only after some time (ca. 65 min for HS1 and 200 min for AHS1), both samples can be equally well described by either a Furukawa function with $\delta = 6$ and Derber's function. The pronounced minimum of the Derber function, however, is never observed: even though HS1 at 65 min displays some indication of an oscillation, it is of too small frequency. The minimum occurs, because the form factor of only one single object (crystal + depletion zone) is calculated, which neglects the polydispersity of the objects in the experiment due to continued nucleation. We anticipate, that this minimum will be smeared out, when the distribution of sizes is taken into account. As time progresses the mid-$Q$-region $1.2 < Q < 2.4$ deviates towards the off-critical Furukawa function ($\delta = 4$), and especially sample AHS1 shows a rather good agreement near the end of the main crystallization process ($t \approx t_{cross}$). Interestingly, for later stages, Furukawa functions of ever decreasing fractal dimension appear to be much closer to our data at large q than throughout stage (B). In general, however, none of the expected shapes is met throughout a complete time stage. Also other types of scaling relations have been suggested [33, 42], which are based on $Q^* = q / q_{1/2}(t)$, where $q_{1/2}$ is the always accessible large-q value corresponding to half the maximum intensity. We tried also these, but could not improve the agreement any further. We have to conclude that the peaked signals evade a scaling analysis due to their continuously changing shape.

In our view, this disagreement contains some physical argument. If our crystallization scenarios were spinodal decomposition like, we would expect the signal shape to remain fixed and the integrated intensity to increase also during stage (C) which, however, is not the case. The assumption of a mass conserving crystal plus depletion zone type SALS object leading to form factor scattering with two independently evolving length scales, however, can in principle explain our observations. In particular, the SALS signals should in this case resemble a critical quench Furukawa function with $\delta = 6$ at some stage, when the fraction of these 3D objects equals the fraction of remaining melt. Moreover, its large q-decay should resemble an off-critical quench Furukawa function of $\delta = 4$ close to $t_{cross}$ when the system comprises well defined crystals embedded in a mesh of 3D melt regions which are still slightly super-saturated. Both indeed is observed in HS1, AHS1 and a number of other samples prepared at coexistence conditions. Since the micro-structure of this scenario is equivalent to that of a wet foam [94] we would further expect $\delta$ to drop as the foam drains and forms facet like 2D grain boundaries and finally 1D plateau borders. Also this development towards lower $\delta$ is seen in our signals. Our observation therefore suggest the interpretation that form factor scattering is at the origin of our signals for samples at coexistence rather than a spinodal decomposition scenario. For sure, further development of the model functions is needed to proceed to a quantitative modelling. None of the here used models considers a size distribution of the SALS objects, neither late stage overlap of depletion zones nor changes of the crystallite density with time are addressed as yet.

HS1 and AHS1 are representative for samples with no or little depletion attraction. AHS2 with its elevated polymer concentration is representative for several samples with strong attractive interactions present. As can be seen in Figure 7g for such samples the SALS peaks are much less pronounced and appear to be located at much smaller q than before. Moreover, no scaling of the SALS signals is observed in Fig. 10a, b. Further, the observed shapes in general do not comply with any of the here available model functions. Only for a limited time interval a Furukawa function with $\delta = 2$ shows some resemblance to the measured data. In fact, the intensity increase and subsequent decrease seems to be restricted to $q < 0.025$ μm$^{-1}$ for times $t < 2000$ min. A noticeable intensity increase for $q > 0.025$ μm$^{-1}$ is observed only for larger times, when the low q intensity already has significantly decreased. Then the signal bears some resemblance to the signals of HS2. Finally, by contrast to the data for HS2 (c.f. Fig. 7), HS1 and AHS1, the large q slope is ill defined and changes irregularly.

Similar observations of a loss of scaling upon increasing the attractive strength have been made already early [49]. For a system of sterically stabilized silica spheres in density matching organic solvent the authors noted a pronounced intensity increase at very low q relative to a much weaker intensity increase at intermediate q. Therefore, for weaker attractive strength the data could be well described by Furukawa-fits, while this did not apply to more strongly interacting samples. Other authors, e.g. [95], have observed a strong increase of low q-scattering intensity upon gel formation. From parallel microscopic investigation, they inferred this increase to correspond to the formation of few, sample spanning thread-like clusters of large size and discriminates such samples from systems with homogeneously distributed small clusters or cluster fluids [96]. In the latter case only moderate increases of the scattered intensity at the lowest q is observed. Together these observations indicate, that the crystallization scenario for strongly attractive systems may differ considerably from those at no or only weak attraction. A possible candidate for the signal origin seems to be the formation of large, gelled threads of compressed fluid or crystalline structure. Here direct microscopic investigations may help to further clarify the situation.

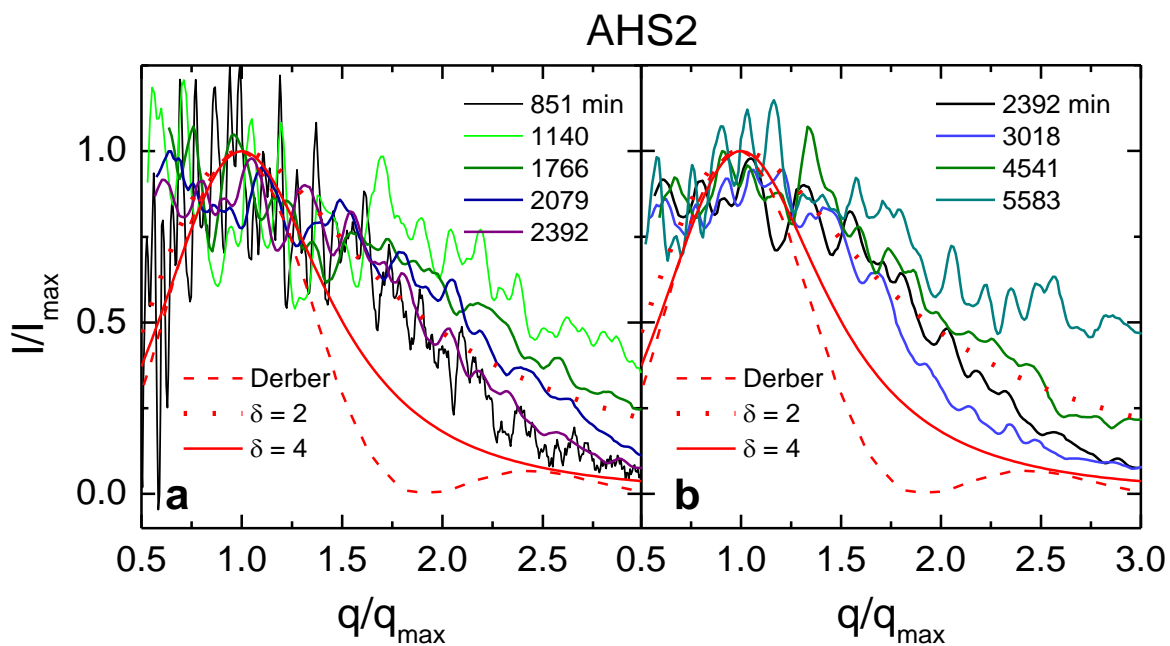

*Fig. 10: Rescaled SALS data of sample AHS2 at selected times during stage B (a) and stage C (b). The data follow no scaling behaviour at all and can only be described by a Furukawa function with $\delta = 2$ at certain times. For comparison with the previously discussed measurements we added Furukawa functions for $\delta = 2$ and 4 and Derber's form factor.*

### iv) Samples with late stage SALS signals (AHS3)

For all micro-gel samples discussed above, the SALS signal shows a significant increase in integrated intensity already during the main crystallization stage (B'). By contrast, AHS3 shows no change in SALS intensity until long after $\tau_{cross}$. Neither a peaked nor an unpeaked signal (as for HS2) could be discriminated during stage (B) due to the much larger small q-background for these signals recorded with the previous machine. The deviating behaviour is clearly seen in the right column of Fig. 6. Further, this short range attractive sample of low attraction strength reveals a remarkably nice off-critical quench Furukawa scaling, as can be seen in Figs. 12a, b. Here our data were background corrected by an early fluid spectrum and Fig. 11 shows good agreement with Furukawa functions of $\delta \approx 2.9$ for all times $t > 400$ min. A slight quantitative discrepancy exists to the data published in [44], where these were corrected by the background after complete crystallization. There, detailed analysis of the fitted Furukawa functions yielded a decrease in $\delta$ from values around 3.2 for $t < 400$ min to values around 2.7 for $t > 800$ min. Additional microscopic investigations on this sample revealed the formation of fluid regions embedding crystallites during the main crystallization stage. Upon successive crystal density relaxation and further coarsening these initially 3-dimensional grain boundaries slowly coarsened to form plate-like and finally rod-shaped regions accommodating the remaining 5% of fluid [43, 44]. Consequently, the explanation of the SALS origin here is based on a change of micro-structure. It involves the narrowing of the crystallite distribution towards a near monodisperse distribution upon coarsening with a conserved order parameter polymer density. At the same time the distance between opposing grain boundaries – closely connected to the increasing average crystallite size – emerged as the characteristic length scale, giving rise to a small-q structure factor.

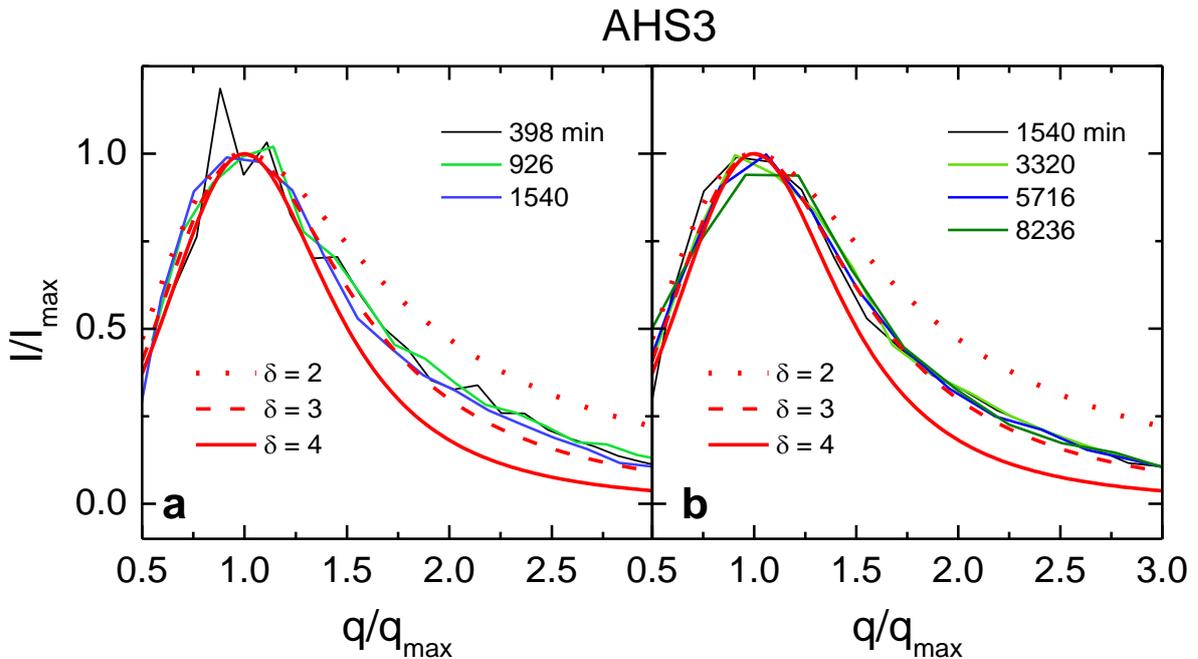

*Fig. 11: Rescaled SALS data of sample AHS3. The data show Furukawa scaling with $\delta = 3$. For comparison we added Furukawa functions with $\delta = 2$ and 4 as well.*

Late stage signals have been observed in the literature before, but have not been interpreted this way. He et al. [39] as well as Ackerson and Schätzel [88] observed signals which showed two different scaling regimes, one during the main crystallization stage and a second, corresponding to much shorter length scales at late stages. Their interpretation involved a crystallite break-up upon intersection and impingement – a scenario actually not supported by microscopic investigation. An interpretation in terms of a superposition of a depletion zone scattering signal, visible already at early times but fading past $\tau_{cross}$ and a late stage signal caused by the emergence of a characteristic length scale during coarsening may provide an interesting alternative.

### v) Correlations of crystallization scenarios and characteristic SALS signal types

The present study revealed a rich variety of different SALS signals and SALS signal evolutions during the crystallization of colloidal HS suspensions. The gauging approach taken here for the first time facilitated extracting qualitative information from the measured SALS signals and in several cases answering questions concerning their origin, the scattering mechanism involved and ultimately also of the crystallite environment and the sample micro-structure evolution in time. In particular we observe that for each crystallization scenarios a typical scattering mechanism is observed.

Crystallization kinetics of fully crystallizing samples like HS2 have so far been investigated by either BS or extinction experiments [93]. In the present study their unpeaked SALS signal shape suggested modelling in terms of a superposition of form factors of polydisperse discs. Different to most examples of amplitude contrast in microscopy, here the change in transmission is not due to absorption, but to the different BS signal for crystal and fluid. The integrated intensity of BS and SALS signals shows a strict temporal correlation in their evolution. For the example given we could obtain a decent signal already early during the precursor stage. To the best of our knowledge, this is the first demonstration of a SALS signal due to amplitude contrast for a crystallizing colloidal suspension. Since the measurements do not need a correction for the remaining fluid, we anticipate that for future experiments this information may provide a useful cross-check for the crystallinity obtained in the BS regime.

In the case of peaked SALS signals we regularly observe an increase of their integrated intensity during the conversion stage, but a subsequent decrease, beginning towards the end of conversion and continuing way into the coarsening stage. This observation excludes spinodal decomposition like scattering with fixed object and repeat length scales which continuously increase with time.

Rather, our comparison to several model functions favour form factor scattering as origin of the SALS signal in these cases. Correlation of the increase of the SALS signal to the growth stage inferred from BS and the existence of a peak here yields the additional constraint, that we have mass conservation between a compressed crystal and its surrounding depletion zone, both growing into the undisturbed melt. As the crystal density relaxes from the initially highly compressed precursor state to the final density close to the melting density, we expect the shape to undergo subtle changes. The change in shape and intensity is expected to be even more pronounced upon the overlap of depletion zones. The observed SALS signal evolution follows these expectations from the depletion zone model. In particular, the initial period of SALS signal strength increase obeys some scaling which is lost later during the growth stage. The SALS signal then weakens and ultimately vanishes. One then is left with the much weaker amplitude scattering only. Unfortunately, Fresnel diffraction fine structure and even more so the polydispersity of the growing objects smears out the minima expected in some of the tested theoretical models and no quantitative analysis could be performed to obtain e.g. depletion zone sizes. Therefore, here our evaluation remained qualitative. We could not detect any depletion zone formation during the precursor stage. Identification of the onset of a peaked signal with the onset of depletion zone formation sometime after first crystals form from precursors allowed to discriminate the stages B and B´ during the growth stage. Future experiments may pay attention also to a characteristic time of depletion zone overlap.

An important role in such experiments may be taken by the scaling properties, as loss of scaling denotes changes in the shape and/or density distribution of the SALS scattering object. Here, however, theoretical expressions are desired, which better match observed signal shapes. Their development should be assisted by additional microscopic investigations to characterize the density distribution within the depletion zones quantitatively.

Proof of principle of the usefulness of parallel investigation by spatially resolved microscopy could be given for another class of SALS signals originating from an additional micro-structural length scale. This length scale is a characteristic distance between grain boundaries, which for thin boundaries roughly coincides with the crystallite size [43, 44]. Such an inter-crystallite ordering is observable only during the late stages of coarsening. Therefore the corresponding structure factor scattering also is observed only way after the end of the main crystallization stage in HS3. Interestingly, in this case the data very closely follow a Furukawa scaling law. It will be interesting to (re-)investigate other samples of HS to see, whether their late stage signals have the same similar origin, if ordering during coarsening is a general phenomenon and whether it is restricted to the upper end of the coexistence range.

Parallel microscopic investigations are anticipated to be particularly useful for samples with strongly attractive interactions. Also there we have observed peaked SALS signals seemingly forming a systematically different class, but could not conclusively determine the underlying scattering mechanism. Also a further extension of the time window covered, e.g. by in-situ ultrasound melting of the mounted samples may in future allow accessing also the earliest stages in such quickly transforming samples and aid the identification of the involved scattering mechanisms.

## V   CONCLUSION

The present investigation of colloidal crystallization revealed a rich variety of qualitatively different SALS signals correlating with different crystallization scenarios and/or different micro-structures for systems undergoing a first order phase transition from a metastable melt to a crystalline state. Their discrimination and interpretation was based on the combined use of SALS to study the evolution of samples on the micro-structural length scale and of BS to obtain a temporal reference from the known crystallization scenario on the structural length scale. We compared to a few theoretical suggestions available from literature to find in most cases a qualitative agreement sufficient to discriminate different SALS signal origins. We demonstrated that SALS signals in HS crystallization may result from both form factor scattering of individual objects and structure factor scattering from objects with characteristic distance. Further scattering contrast may be provides either by phase contrast or amplitude contrast. As both wall and bulk crystals contribute, and BS is always present, different scattering mechanisms in general superimpose. Within suitable models for the shape of the signal, SALS data may in principle be further quantified in terms of characteristic length scales, dimensionality and/or kinetic exponents.

We demonstrated that the parallel investigation of crystallizing colloidal suspensions with small angle light scattering and Bragg scattering yields a wealth of additional complementary information on system structure, micro-structure and crystal environments. On one side the applied temporal gauging derived from the BS, allowed an in depth discussion of the SALS signals. On the other side, the identification of the involved scattering mechanism and the interpretation of the SALS signals revealed important kinetic details not accessible from BS alone.

The present study used samples of rather different crystallization scenarios to obtain a survey on the wealth of possible SALS signals. In future, we will use this experiment in combination with microscopy to further study different volume fractions of a single system class across the coexistence region and above melting. From such data we anticipate to derive a quantitative model for the density distribution within and around the growing crystallites and calculate the corresponding time dependence of the SALS pattern. Further, the approach taken is not restricted to hard spheres, but easily adjustable for other systems. Most promising seem to be investigations of crystallizing charged sphere suspensions and eutectic mixtures undergoing phase separation upon crystallization.

**ACKNOWLEDGEMENTS**

It is a pleasure to thank H. Moschallski and O.Thorwarth from the group of E. Bartsch, Freiburg, for particle synthesis and characterization measurements. We further thank S. Iacopini and J. Viitala for technical support. Financial support of the DFG (Scho1054/3; Pa459/13, 16 and 17; 13). The EU MCRT-Network CT-2003-504 712, and the Graduate School of Excellence, MAINZ, are gratefully acknowledged.

**APPENDIX A: Light scattering setup**

The light scattering machine, used in this work, is based on a design introduced by Heymann et al. [58]. It has been improved in angular and temporal resolution by replacing the former diode arrays with linear CCD detectors.

The sample is illuminated using a widened laser beam (7 mW He-Ne, $\lambda_0 = 632.8$ nm, NEC, UK) of 8 mm diameter. The light passes a coupled polariser and $\lambda/4$-plate unit which by rotation allows to adjust the illumination intensity and ensures circular polarization. The beam is made slightly convergent, such that with sample stage mounted, it focuses exactly at the distance of the SALS detection unit.

The sample stage is home-built with optical parts custom made from glass with $n_{D,Glass}$ = 1.602 (Berliner Glas, Berlin, Germany). These comprise of a rectangular index match bath containing 1-EN ($n_{D,1EN}$ = 1.606) and a hemispherical glass lens with a flattened apex. Samples of average index of refraction of $n_{D,sample} \approx$ 1.604 are contained in rectangular quartz cuvettes (30 x 10 x 5 mm$^3$, Starna, Germany) and placed inside the match bath. Both stage and sample holder allow for tilt and rotation to facilitate precise alignment of all optical surfaces normal to the optical axis. This construction and choice of materials minimizes refraction effects inside the bath (except for a minimal beam displacement at the cuvette walls) and renders the detected intensity distribution practically free of parasitic reflections.

The hemispherical lens has a diameter of 110 mm and hence a focal length of f = 150 mm. It is carefully centred on the optical axis using a µm-gauge mounted to the rotating rails. Bragg scattered light passes normal to the lens-air interface and is focused onto a spherical surface, in which the detector units are placed (Fig. 12a). This plane corresponds to the back focal plane of a microscope and thus contains the distribution of intensity in reciprocal space independent of the real space distribution of the individual scattering objects. To detect the complete 2D scattering pattern, six linear CCD arrays (LARRY, LOT-Oriel, Germany) are mounted on four curved rails and adjusted by µm translation stages. Their angular detection ranges slightly overlap. This greatly enhances the performance as compared to the single rail equipped with a single diode array detector used in the previous machine [33, 35].

Angular calibration of the CCD units was performed using high quality optical gratings [Edmund Optics, Germany] inserted into the index matching bath. We observed the first six interference maxima of a 10±0.5 lines/mm grating for the SALS regime and the maxima of order 11 to 34 of a 70±3.5 lines/mm grating for the BS regime. Fig. 12b displays the observable maxima in both regimes. Comparison to the calculated positions of the maxima allowed to allocate the pixel positions to better than $(3\times10^{-4})°$ for the SALS regime and better than $(5\times10^{-3})°$ for the BS regime, respectively. Apart from the manufacturing tolerance, the main uncertainty stems from the noise in the region of maximum intensity. The absolute uncertainty increases with increasing angle from $(4\times10^{-4})°$ for the smallest SALS angle to 0.3° for the largest BS angle. The relative uncertainty, however is roughly constant at $\Delta\Theta = 10^{-3}\,\Theta$ and thus about an order of magnitude better than for the previous instrument. The signals of the individual detectors are corrected for sensitivity differences in different pixels and are then merged to yield a single continuous signal I(q) covering an angular range of $\Theta$ = (16.450 – 73.388)°, equivalent to q = (4.562 – 19.057) µm$^{-1}$. Each CCD array hosts 2040 pixels of 14 µm width. Our nominal angular resolution of 181.2 pixels/° (or $\Delta q$ = 1.41 10$^{-3}$ µm$^{-1}$ per pixel) for the arrays mounted at the BS focus surpasses that of the previous diode array detector units [58] by about two orders of magnitude.

Fig. 12a shows that both scattered and unscattered light pass the central part of the lens. Refraction at the planar lens-air interface enlarges the detectable scattering angles $\theta_{SALS}$ and at the same time increases the convergence of the illuminating beam again. Its focus on the optical axis defines the SALS detection plane to be at $d_{SALS}$ = 146.2 cm past the sample centre. This construction allows a convenient separation of the off-axis SALS signal and the on-axis unscattered light. With the present choice of $d_{SALS}$ we already obtain a high angular resolution for our signal, while the mechanical strength of the set-up is still retained.

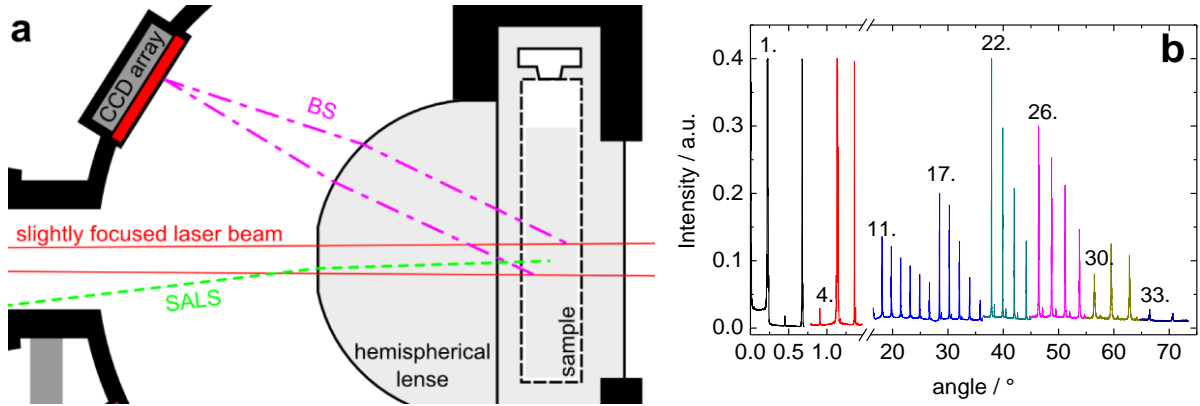

*Fig. 12: The close-up a) shows the slightly focused laser beam (solid line) and two typical optical paths for the light scattered into the BS regime (dash dotted lines) and the SALS regime (dashed line). Note that the former is focused by the hemispherical lens onto the detector, while the latter is made more divergent by refraction at the planar rear window of the lens. b) Intensity scattered off a 10 lines per mm (70 lines per mm) grating to calibrate the range of accessible azimuth angles for SALS (BS) measurements. Numbers indicate the nth maximum, each being the lowest order detected by a detector 1 and 2 for SALS (3-7 for BS).*

For SALS detection, two CCD arrays are mounted normal to the optical axis on the two sides of a planar rail and shielded against stray light by a metal box. Also the two SALS detection ranges are made to overlap in order to merge signals. The current construction covers an angular range of $\Theta_{SALS}$ = (0.042 - 1.442)° with a resolution of 2864 pixel/°, corresponding to $q_{SALS}$ = (0.012 - 0.401) µm$^{-1}$ with $\Delta q_{SALS}$ = 9.7 10$^{-5}$ µm$^{-1}$. Compared to the previous diode array SALS instrument [58], this roughly halves the covered q-range and shifts it to smaller values, while at the same time the nominal q-resolution is increased by nearly two orders of magnitude. All 8 detectors are read out in parallel and their analog read out is converted to serial digital signal by an AD-converter (SAM8, LOT-Oriel, Germany) and stored to a personal computer for later evaluation.

Each full forward-backward rotation of the two coupled detectors defines a single measurement run. During forward rotation, we acquire 35 slices of the Debye-Scherer-cone with all eight detectors. Each slice consists of an average over five intensity measurements of at least 10 ms duration. For the backward rotation, we acquire 6 control slices, each consisting of an average over 90 intensity measurements with the first and the fourth detector. With mechanical motion in each direction taking some 30s, the minimum repetition time is about 80s. This temporal resolution is less by about a factor of two as compared to the previously used instrument [58].

To verify the static operation capabilities of the SALS system we measure the form factor of large Polystyrene microspheres (*BS-Partikel, Wiesbaden, Germany*) shown in Fig. 14a and b. Their manufacturer given nominal diameter is 2a = (38.0 ± 0.4) μm. A suspension of $\Phi \approx 0.02$ was prepared by adding 1EN to the stock powder. We measured a single run SALS signal and corrected for solvent background scattering, the Gaussian of the laser beam and parasitic stray light by subtracting a measurement on the pure solvent: $I(q) = I(q)_{Suspension} - I(q)_{Solvent}$. Note that this produces a comparably large scatter at low q. For comparison, we also calculated an estimate of a theoretical form factor. Observed in bright field under the microscope (Fig. 14b), each particle shows pronounced airy fringing typical for a core-shell particle with the inner dark fringe marking the location of the core-shell boundary. From the dark fringe position we obtain the core radius giving the dominant scattering contribution. We obtain an average core radius of $a_{core}$ = (14.95 ± 0.7) μm from which a theoretical particle form factor was calculated using the MIETAB [97] software code for monodisperse particles. The result is compared in Fig. 14a to the experimental data. Here the experimental intensity was scaled to meet the expectation best up to the first minimum. There is an increasing deviation at large angles which is attributed to a small amount of multiple scattering, not accounted for in the MIETAB calculation. Otherwise our experimental data meets the expected form factor remarkably well and in particular the position of the minimum is exactly met. This clearly demonstrates that quantitative scattering measurements with good statistics can be performed with the present combined setup. We may thus detect density fluctuations on scales between about 15 and 500 μm for the SALS regime and between about 0.31 and 1.26 μm in the BS regime with a temporal resolution of about one minute.

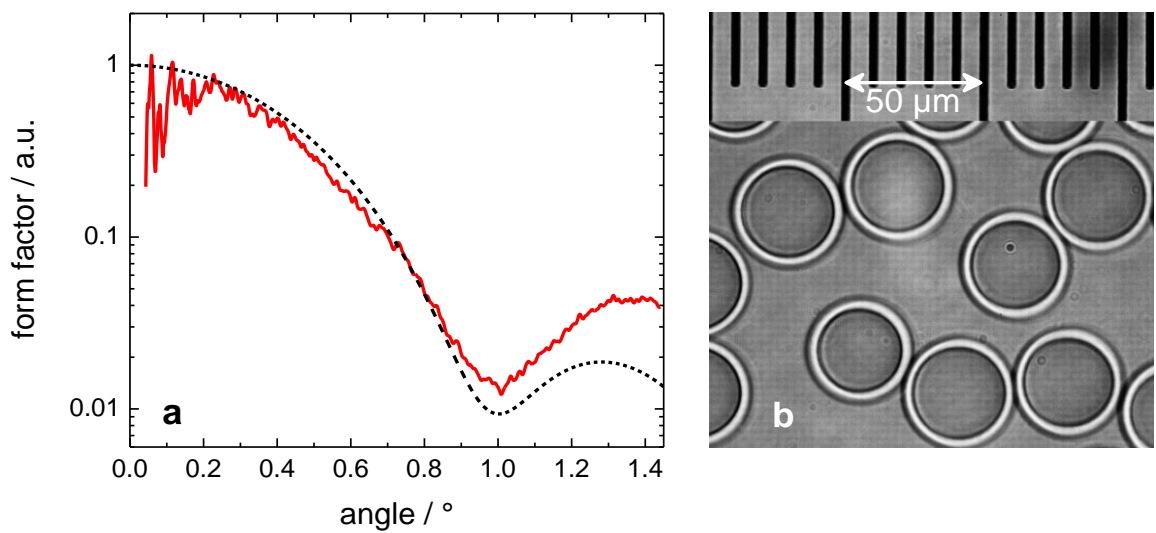

*Fig. 13: a) The form factor measurement of large PS particles (red, solid curve) in comparison with a MIETAB form factor calculation for a radius of a = (14.95 ± 0.7) μm (black dotted line) measured by microscopy. b) typical micrograph of the large PS particles revealing a pronounced fringing typical for core shell particles with thick core of large difference in refractive index to that of the solvent.*

**REFERENCES**

1   P. N. Pusey, W. van Megen, *Phase Behaviour in Concentrated Suspensions of Nearly Hard Colloidal Spheres*, Nature **320**, 340-342 (1986).

2   P. Bartlett, W. v. Megen, Granular Matter, *The Physics of Hard-Sphere Colloids*, edited by A. Mehta (Springer, New York, 1994), p. 195-257.

3   W. van Megen, *Crystallization and the glass transition in suspensions of hard colloidal spheres,* Transport Theory and Statistical Phys. **24**, 1017 – 1051 (1995).

4   B. J. Ackerson (ed.), special issue *Phase transitions in Colloidal suspensions*, Phase Transitions **21,** (2-4), 73 - 249 (1990).

5   T. Palberg, *Crystallisation kinetics of repulsive colloidal spheres,* J. Phys.: Condens. Matter **11**, R323 - R360 (1999).

6   W. C. K. Poon, *The physics of a model colloid–polymer mixture*, J. Phys.: Condens. Matter **14,** R859 (2002).

7   S.-H. Chen, W.-R. Chen, F. Mallemace, *The glass to glass transition and its endpoint in a copolymer micellar system*, Science **300**, 619-622 (2003).

8   V. J. Anderson, H. N. W. Lekkerkerker, *Insights into phase transition kinetics from colloidal science*, Nature **416**, 811 (2002).

9   E. Bartsch, T. Eckert, C. Pies, H. Sillescu, *The effect of free polymer on the glass transition dynamics of microgel colloids*, J. Non-Cryst. Solids **307 – 310**, 802 (2002).

10  R. P. Sear, *Nucleation: theory and applications to protein solutions and colloidal suspensions,* J. Phys.: Condens. Matter **19**, 033101 (2007).

11  T. Palberg, *Crystallization kinetics of colloidal model suspensions: recent achievements and new perspectives*, J. Phys.: Condens. Matter **26**, 333101 (2014).

12  H. J. Schöpe, G. Bryant, W. van Megen, *Two step crystallization kinetics in colloidal hard spheres,* Phys. Rev. Lett. **96**, 175701 (2006).

13  T. Schilling, H. J. Schöpe, M. Oettel, G. Opletal, I. Snook, *Precursor-Mediated Crystallization Process in Suspensions of Hard Spheres,* Phys. Rev. Lett. **105**, 025701 (2010).

14  T. Kawasaki, H. Tanaka, *Formation of a crystal nucleus from liquid,* Proc. Nat. Acad. Sci. **107,** 14036-14041 (2010).

15  J. Russo, H. Tanaka, *The microscopic pathway to crystallization in undercooled fluids,* Sci. Rep. **2**, 505 (2012).

16  P. Tan, N. Xu, L. Xu, *Visualizing kinetic pathways of homogeneous nucleation in colloidal crystallization,* Nature Phys. **10**, 73–79 (2014).

17  K. Kratzer, A. Arnold, *Two-stage crystallization of charged colloids under low supersaturation conditions* Soft Matter **11**, 2174-2182 (2015).

18  H. J. Schöpe, T. Palberg, *Frustration of structural fluctuations upon equilibration of shear melts*, J. Non-Cryst. Mater **307-310**, 613 – 622 (2002).

19  M. Franke, S. Golde, and H. J. Schöpe, *Solidification of a colloidal hard sphere like model system approaching and crossing the glass transition*, Soft Matter **10** 5380-5389 (2014).

20  U. Gasser, *Crystallization in three- and two-dimensional colloidal suspensions,* J. Phys.: Condens. Matter **21**, 203101 (2009).

21  N. J. Lorenz, H. J. Schöpe, H. Reiber, T. Palberg, P. Wette, I. Klassen, D. Holland-Moritz, D. Herlach and T. Okubo, *Phase behaviour of deionized binary mixtures of charged colloidal spheres*, J. Phys.: Condens. Matter **21,** 464116 (2009).

22  S. R. Ganagalla and S. N. Punnathanam, *Free energy barriers for homogeneous crystal nucleation in a eutectic system of binary hard spheres,* J. Chem. Phys. **138**, 174503 (2013).

23  M. Leocmach, C. P. Royall, H. Tanaka, *Novel zone formation due to interplay between sedimentation and phase ordering,* Europhys. Lett. **89**, 38006 (2010).

24  A. Kozina, P. Diaz-Leyva, C. Friedrich, E. Bartsch, *Structural and dynamical evolution of colloid-polymer mixtures on crossing glass and gel transition as seen by optical microrheology and mechanical bulk rheology*, Soft Matter **8**, 627-630 (2012).

25  J. Nozawa, S. Uda, Y. Naradate, H. Koizumi, K. Fujiwara, A. Toyotama, and J. Yamanaka, *Impurity Partitioning During Colloidal Crystallization,* J. Phys. Chem. B **117**, 5289−5295 (2013).

26   A. Kozina, P. Diaz-Leyva, T. Palberg, E. Bartsch, *Crystallization kinetics of colloidal binary mixtures with depletion attraction*, Soft Matter **10**, 9523 (2014).

27   A. Engelbrecht, H. J. Schöpe, *Drastic Variation of the Microstructure Formation in a Charged Sphere Colloidal Model System by Adding Merely Tiny Amounts of Larger Particles*, Crystal Growth and Design **10**, 2258 – 2266 (2010).

28   M. Franke, A. Lederer, H. J. Schöpe, *Heterogeneous and homogeneous crystal nucleation in colloidal hard-sphere like microgels at low metastabilities*, Soft Matter **7**, 11267-11274 (2011).

29   K. Sandomirski, S. Walta, J. Dubbert, E. Allahyarov, A.B. Schofield, H. Löwen, W. Richtering, S.U. Egelhaaf, *Heterogeneous crystallization of hard and soft spheres near flat and curved walls,* Euro. Phys. J. Special Topics **223**, 439-454 (2014).

30   T. Okubo, *Giant colloidal single crystals of polystyrene and silica spheres in de-ionized suspensions,* Langmuir **10**, 1695 – 1702 (1994).

31   T. Palberg, M. R. Maaroufi, A. Stipp and H. J. Schöpe, *Micro-structure evolution of wall based crystals after casting of model suspensions as obtained from Bragg microscopy*, J. Chem. Phys. **137**, 094906 (2012).

32   M. Shinohara, A. Toyotama, M. Suzuki, Y. Sugao, T. Okuzono, F. Uchida, and J. Yamanaka, *Recrystallization and Zone Melting of Charged Colloids by Thermally Induced Crystallization,* Langmuir **29**, 9668−9676 (2013).

33   K. Schätzel, B. J. Ackerson, *Observation of Density Fluctuations During Crystallization,* Phys. Rev. Lett. **68**, 337 (1992).

34   M. D. Elliot, W. C. K. Poon, *Conventional optical microscopy of colloidal suspensions,* Adv. Colloid Interface Sci. **92**, 133 – 194 (2001).

35   K. Schätzel, B. J. Ackerson, *Crystallization of Hard Sphere Colloids,* Phys. Scripta **T49**, 70-73 (1993).

36   K. Schätzel, B. J. Ackerson, *Density Fluctuations during Crystallization of Colloids,* Phys. Rev. E **48**, 3766 (1993).

37   B. J. Ackerson, K. Schätzel, *Classical growth of hard-sphere colloidal crystals,* Phys. Rev. E **52**, 6448–6460 (1995).

38   S. Derber, T. Palberg, K. Schätzel, J. Vogel, *Growth of a colloidal crystallite of hard spheres,* Physica A **235**, 204-215 (1997).

39   Y. He, B. J. Ackerson, W. van Megen, S. M. Underwood, K. Schätzel, *Dynamics of crystallization in hard-sphere suspensions,* Phys. Rev. E **54**, 5 (1996).

40   Y. He, B. J. Ackerson, *Crystallization of hard spheres and turbidity,* Physica A **235**, 194 – 203 (1997).

41   A. Heymann, A. Stipp, Chr. Sinn, T. Palberg, *Observation of oriented close-packed lattice planes in polycrystalline hard-sphere solids,* J. Coll. Interface Sci. **207**, 119 - 127 (1998).

42   C. Sinn, A. Heymann, A. Stipp, T. Palberg, *Solidification kinetics of hard-sphere colloidal suspensions*, Prog. Colloid Polym. Sci. **118**, 266 – 275 (2001).

43   T. Palberg, A. Stipp, E. Bartsch, *Unusual Crystallization Kinetics in a Hard Sphere Colloid-Polymer Mixture*, Phys. Rev. Lett. **102**, 038302 (2009).

44   A. Stipp, H.-J. Schöpe, T. Palberg, T. Eckert, R. Biehl, E. Bartsch, *Optical experiments on a crystallizing hard-sphere–polymer mixture at coexistence*, Phys. Rev. E **81**, 051401 (2010).

45   D. Ashnagi, M. Carpineti, M. Giglio, and A. Vailati, *Small angle light scattering studies concerning aggregation processes,* Curr. Opinion Colloid Interface Sci. **2**, 246-250 (1997).

46   M. Carpineti and M. Giglio, *Spinodal-type dynamics in fractal aggregation of colloidal clusters,* Phys. Rev. Lett. **68**, 3327-3330 (1992).

47   M. Carpineti, M. Giglio, and V. Degiorgio, *Mass conservation and anticorrelation effects in the colloidal aggregation of dense solutions,* Phys. Rev. E. **51**, 590-596 (1995).

48   A. E. Bailey et al., *Spinodal Decomposition in a Model Colloid-Polymer Mixture in Microgravity*, Phys. Rev. Lett. **99**, 205701 (2007).

49   N. A. M. Verhaegh, J. S. van Duijneveldt, J. K.G. Dhont, H. N.W. Lekkerkerker, *Fluid-fluid phase separation in colloid-polymer mixtures studied with small angle light scattering and light microscopy*, Physica A **230**, 3–4, 409–436 (1996).

50   H. Furukawa, *Dynamic scaling theory for phase-separating unmixing mixtures: Growth rates for droplets and scaling properties of autocorrelation functions,* Physica A **123**, 497-515 (1984).

51  I. M. Lifshitz and V.V. Slyozov, *The kinetics of precipitation from supersaturated solid solutions*, J. Phys. Chem. Solids **19**, 35 (1961).

52  J. D. Gunton, M. S. Miguel, and P. S. Sahni, *Phase Transitions and Critical Phenomena* Vol. 8, edited by C. Domb and J. L. Lebowitz (Academic, New York, 1983).

53  A. J. Bray, *Theory of phase ordering kinetics,* Adv. Phys. **43**, 357 (1994).

54  K. Binder, *Theory of first-order phase transitions,* Rep. Prog. Phys., **50**, 783 (1987).

55  P. J. Lu, E. Zaccarelli, F. Ciulla, A. B. Schofield, F. Sciortino, D. A. Weitz, *Gelation of particles with short-range attraction,* Nature **453**, 499 - 503 (2008).

56  A. Weiss, K. D. Hörner, and M. Ballauff, *Analysis of Attractive Interactions between Latex Particles in the Presence of Nonadsorbing Polymers by Turbidimetry,* J. Colloid Interface Sci. **213**, 417-425 (1999).

57  C. M. Sorensen, *Light Scattering by Fractal Aggregates: A Review,* Aerosol Sci. Tech. **35**, 648-687 (2001).

58  A. Heymann, A. Stipp, K. Schätzel, *Scaling in colloidal crystallization,* Il Nouvo Cimento D **16**, 1149 – 1157 (1994).

59  V. C. Martelozzo, A. B. Schofield, W. C. K. Poon, P. N. Pusey, *Structural aging of crystals of hard-sphere colloids,* Phys. Rev. E **66**, 021408 (2002).

60  T. Zykova-Timan, J. Horbach, and K. Binder, *Monte Carlo simulations of the solid-liquid transition in hard spheres and colloid-polymer mixtures,* J. Chem. Phys. **133**, 014705 (2010).

61  P. G. Bolhuis, D. A. Kofke, *Monte carlo study of freezing of polydisperse hard spheres,* Phys. Rev. E **54**, 634 (1996).

62  M. Fasolo and P. Sollich, *Fractionation effects in phase equilibria of polydisperse hard-sphere colloids,* Phys. Rev. E **70**, 041410 (2004).

63  H. N. W. Lekkerkerker, W. C. K. Poon, P. N. Pusey, A. Stroobants, P. B. Warren, *Phase Behaviour of Colloid + Polymer Mixtures*, Europhys. Lett. **20**, 559 (1992).

64  M. Fasolo and P. Sollich, *Effects of colloid polydispersity on the phase behavior of colloid-polymer mixtures,* J. Chem. Phys. **122**, 07904 (2005).

65  R. Beyer, S. Iacopini, T. Palberg and H. J. Schöpe, *Polymer induced changes of the crystallization scenario in suspensions of hard sphere like microgel particles*, J. Chem. Phys. **136**, 234906 (2012).

66  A. Kozina, P. Díaz-Leyva, T. Palberg, E. Bartsch, *Crystallization kinetics of colloidal binary mixtures with depletion attraction*, Soft Matter **10**, 9523-9533 (2014).

67  M. Dijkstra, J. M. Brader, R. Evans, *Phase behaviour and structure of model colloid–polymer mixtures,* J. Phys.: Condens. Matter **11**, 10079-10106 (1999).

68  B. J. Ackerson, S. E. Paulin, B. Johnson, W. van Megen, S. Underwood, *Crystallization by settling in suspensions of hard spheres,* Phys. Rev. E **59**, 6903 (1999).

69  Y. He, B. Olivier, B. J. Ackerson, *Morphology of crystals made of hard spheres,* Langmuir **13**, 1408 – 1412 (1997).

70  W. B. Russel, P. M. Chaikin, J. Zhu, W. V. Meyer, R. Rogers, *Dendritic growth of hard sphere crystals,* Langmuir **13**, 3871 (1997).

71  W. Poon, F. Renth, R. M. L. Evans, D. J. Fairhurst, M. E. Cates, and P. N. Pusey, *Colloid-Polymer Mixtures at Triple Coexistence: Kinetic Maps from Free-Energy Landscapes,* Phys. Rev. Lett. **83**, 1239 (1999).

72  S. Buzzaccaro, R. Rusconi, R. Piazza, *"Sticky" Hard Spheres: Equation of State, Phase Diagram, and Metastable Gels,* Phys. Rev. Lett. **99** 098301 (2007).

73  N. Lorenz, H. J. Schöpe, T. Palberg, *Phase behavior of a de-ionized binary mixture of charged spheres in the presence of gravity*, J. Chem. Phys. 131,134501 (2009).

74  J. Russo, A. C. Maggs, D. Bonn, and H. Tanaka, *The interplay of sedimentation and crystallization in hard-sphere suspensions,* Soft Matter 9, 7369-7383 (2013).

75  T. Eckert, E. Bartsch, *Re-entrant glass transition in a colloid-polymer mixture with depletion attractions,* Phys. Rev. Lett **89**, 125701 (2002).

76  T. Eckert, E. Bartsch, *Glass transition dynamics of hard sphere like microgel colloids with short-ranged attractions,* J. Phys.: Condens. Matter **16**, S4937 (2004).

77  H. Senff and W. Richtering, *Temperature sensitive microgel suspensions: Colloidal phase behavior and rheology of soft spheres,* J. Chem. Phys. **111**, 1705 (1999).

78  M. Wiemann, Ph.D. thesis, Freiburg, 2013

79  A. Kozina, Ph.D. thesis, Freiburg, 2009.

80  S. E. Paulin, B. J. Ackerson, *Observation of a phase transition in the sedimentation velocity of hard spheres,* Phys. Rev. Lett. **64**, 2663 – 2666 (1990) and (Erratum) Phys. Rev. Lett. **65**, 668.

81  M. Fasolo, P. Sollich, *Equilibrium phase behaviour of polydisperse hard spheres,* Phys. Rev. Lett. **91**, 068301 (2003).

82  J. L. Harland, W. van Megen, *Crystallization kinetics of suspensions of hard colloidal spheres,* Phys. Rev. E **55**, 3054 – 3067 (1997).

83  L. Fetters, N. Hadjichristidis, J. Lindner, J. Mays, *Molecular Weight Dependence of Hydrodynamic and Thermodynamic Properties for Well‐Defined Linear Polymers in Solution*, J. Phys. Chem. Ref. Data **23**, 619 (1994).

84  P. S. Francis, S. Martin, G. Bryant, W. van Megen, P.A. Wilksch, *A Bragg scattering spectrometer for studying crystallization of colloidal suspensions,* Rev. Sci. Intr. **73**, 3878 – 3884 (2002).

85  P. N. Pusey, W. van Megen, P. Bartlett, B. J. Ackerson, J. G. Rarity, and S. M. Underwood, *Structure of crystals of hard colloidal spheres,* Phys. Rev. Lett. **63**, 2753 (1989).

86  S. Iacopini, T. Palberg, H. J. Schöpe, *Crystallization kinetics of polydisperse hard-sphere-like microgel colloids: Ripening dominated crystal growth above melting*, J. Chem. Phys. **130**, 084502 (2009).

87  Z. Cheng, P. M. Chaikin, J.X. Zhu, W. B. Russel, W. V. Meyer, *Crystallization kinetics of hard spheres in microgravity in the coexistence regime: Interactions between growing crystallites,* Phys. Rev. Lett. **88**, 015501 (2001).

88  B. J. Ackerson, K. Schätzel, *Complex Fluids* edited by L. Garrido (Springer, Heidelberg, 1992) p. 15-32.

89  J.K.G. Dhont, *An introduction to the dynamics of colloids* (Elsevier, Amsterdam, 1996).

90  M. Born, *Optik* (Springer, Berlin, 1972).

91   J. L. Langford, A. J. C. Wilson, *Scherrer after Sixty Years: A Survey and Some New Results in the Determination of Crystallite Size,* J. Appl. Cryst. **11**, 102-113 (1978).

92   R. J. Spry, D. J. Kosan, *Theoretical Analysis of the Crystalline Colloidal Array Filter,* Appl. Spectroscopy **40**, 782-784 (1986).

93   D. J. W. Aastuen, N. A. Clark, J. C. Swindal, C. D. Muzny in [4], pp139

94   W. Drenckhan, D. Langevin, *Monodisperse Foams in one to three dimensions,* Curr. Opn. Colloid Interface Sci. **15**, 341–358 (2010).

95   P. J. Lu, E. Zaccarelli, F. Ciulla, A. B. Schofield, F. Sciortino, D. A. Weitz, *Gelation of particles with short-range attraction,* Nature **453**, 499 - 503 (2008).

96   P. J. Lu, J. C. Conrad, H. M. Wyss, A. B. Schofield, and D.A. Weitz, *Fluids of clusters in attractive colloids,* Phys. Rev. Lett. **96** 028306 (2006).

97   http://amiller.nmsu.edu/mietab.html, March 2013.